\documentclass[11pt]{article}
\usepackage{latexsym,cite,amssymb}
\usepackage{times}

\makeatletter
\@addtoreset{equation}{section} \makeatother

\input amssym.def
\input amssym.tex

\newcommand{\diag}{{\rm diag}}

\newcommand{\half}{{{\textstyle\frac{1}{2}}}}

\newcommand{\be}{\begin{equation} }
\newcommand{\ee}{\end{equation} }
\newcommand{\ba}{\begin{array}}
\newcommand{\ea}{\end{array}}

\newcommand{\SO}{\mbox{SO}}
\newcommand{\SU}{\mbox{SU}}

\def\cC{{\cal C}}
\def\cE{{\cal E}}
\def\cF{{\cal F}}
\def\cJ{{\cal J}}

\def\cJ{{\cal J}}
\def\cM{{\cal M}}
\def\cP{{\cal P}}
\def\cT{{\cal T}}

\def\Tr{{\rm Tr}}
\def\I_N{{1_{\scriptscriptstyle N\times N}}}

\def\sI{{I}}
\def\sJ{{J}}
\def\sK{{K}}
\def\salpha{{\alpha}}

\begin{document}
\begin{titlepage}
\title{\vskip -60pt
\vskip 20pt Classification of the BPS states in Bagger-Lambert Theory\\
~}
\author{\sc Imtak Jeon,${}^{\dagger}~$ Jongwook Kim,${}^{\dagger}~$ Nakwoo Kim,${}^{\ast}~$\\~\\
\sc Sang-Woo Kim${}^{\natural}~$
~and~ Jeong-Hyuck Park${}^{\dagger}$}
\date{}
\maketitle \vspace{-1.0cm}
\begin{center}
~~~\\
${}^{\dagger}$Department of Physics, Sogang University, Seoul 121-742, Korea\\
~{}\\
${}^{\ast}$Department of Physics and Research Institute of Basic Science\\
Kyung Hee University, Seoul 130-701, Korea\\
~{}\\
${}^{\natural}$Center for Quantum Spacetime, Sogang University, Seoul 121-742, Korea\\
~{}\\
%{\small{Electronic correspondence: {{{park@sogang.ac.kr}}}}}
~~~\\
~~~\\
\end{center}
\begin{abstract}
\noindent We classify, in a group theoretical manner, the BPS
configurations in the  multiple M2-brane theory recently proposed by
Bagger and Lambert. We present  three  types of  BPS equations
preserving various fractions of supersymmetries: in the
first type we have constant fields and the interactions are purely
algebraic in nature; in the second type the equations are invariant
under spatial rotation $\SO(2)$, and the fields can be
time-dependent; in the third class the equations are invariant under
boost $\SO(1,1)$ and provide the eleven-dimensional generalizations of
the Nahm equations.  The BPS equations for different number of
supersymmetries exhibit the division algebra structures: octonion,
quarternion or complex.
%%%
%%We discuss some  solutions, including time dependent fuzzy M2 or M5-branes.
%%%
\end{abstract}
%%%
{\small
\begin{flushleft}
%%%
%~~~~~~~~\textit{PACS}: \\
%~~~~~~~~\textit{Keywords}: supersymmetry, M-theory matrix model, time dependent background
\end{flushleft}}
%%%
\thispagestyle{empty}
\end{titlepage}
\newpage

\tableofcontents %%
\baselineskip18pt
%%\begin{document} --> JHEP
%%%
\section{Introduction}
In a series of   recent papers~\cite{BL},  Bagger and Lambert (BL)
have constructed a  three-dimensional, interacting superconformal
gauge theory of multiple  M2-branes. The action is maximally
supersymmetric with 16 ordinary supersymmetries, and it has been
verified that the theory is indeed superconformal with 16 conformal
supercharges in \cite{Bandres:2008vf}. In the quest for the final
form of the theory, as usual,  it was supersymmetry that provided
crucial guiding lights. The work was initiated as an attempt to
incorporate Basu and Harvey's generalized Nahm equation -which was a
proposal to describe M2-branes ending on an
M5-brane~\cite{Basu:2004ed}- in the full supersymmetric M2-brane
action. Their analysis revealed a novel algebraic structure, namely
the  3-algebra, which is also investigated independently by
Gustavsson~\cite{Gustavsson:2007vu}. Since the discovery, the
multiple M2-brane theory of Bagger and Lambert has attracted an
enormous degree of
attention~\cite{Ganor:2006ub,Ho:2007vk,Copland:2007by,Chen:2007ir,
MacConamhna:2007fw,Berman:2007bv,Chen:2007tt,Bandos:2008um,Gustavsson:2008dy,Mukhi:2008ux,
Berman:2008be,VanRaamsdonk:2008ft,Morozov:2008cb,Lambert:2008et,Gran:2008vi,Ho:2008bn,Gomis:2008cv,
Bergshoeff:2008cz,Hosomichi:2008qk,Papadopoulos:2008sk,Gauntlett:2008uf,Shimada:2008xy,Papadopoulos:2008gh,
Ho:2008nn,Gomis:2008uv,Benvenuti:2008bt,Ho:2008ei,Morozov:2008rc,Honma:2008un,Fuji:2008yj,
Ho:2008ve}. One might expect that, given this genuine superconformal
field theory, $\cM$-theory is now about to unveil its mysterious and
fundamental features.

In the present paper, we set out to classify the BPS states, or the
BPS equations of the BL theory using a group theoretical
consideration. Apparently the theory of our interest has the Lorentz
group $\SO(1,2)$ and the R-symmetry group $\SO(8)$. Instead of
providing the full and thorough survey of possible BPS equations, we
focus mainly on two different types of BPS equations with different number
of supersymmetries, and classify them completely.
The first class is
completely Lorentz invariant, and the other is invariant under the
spatial rotation.

In the first type,  the BPS equations  are given
purely in terms of the three-algebra commutators  and independent of the three-dimensional worldvolume
coordinates. Thus the corresponding nontrivial configurations possess infinite energy, typically corresponding to BPS objects of infinite size. Previously known analogous  algebraic  soultions include the longitudinal M5-brane in $\cM$-theory matrix model which is realized in terms of Heisenberg algebra or large $N$ matrices \cite{Banks:1996nn}.

In the other type the equations are  $\SO(2)$ rotation invariant,
and the fields can be {\it time-dependent}.
A technical reason why we focus on the two classes is that in these cases, fully utilizing the $\SO(8)$ triality we are able to classify the BPS equations completely.

In addition to the two classes,  there is another
possibility to obtain third type of BPS equations \textit{via}   simple tensor product. Namely one can obtain various generalizations of the
  Nahm equations   which are invariant under the boost $\SO(1,1)\subset\SO(1,2)$.
  Our BPS equations manifest  the division algebra structures:
  octonion, quarternion or complex. In the paper we will mainly focus
  on the BPS equations themselves. Our results hold for both the finite and infinite dimensional three-algebras. 
  Note however that the Lorentz invariant BPS equations can have nontrivial solutions only for infinite dimensional three-algebras.  The specific solutions and the physical interpretation will be presented in a   separate
   publication \cite{preparation}.

The organization of the present paper is as follows.  Sec.\ref{secPre}
 is for  preliminaries.
 We first discuss the general features of the `supersymmetric projection matrices' and
 review how to derive the corresponding  BPS equations for a given projection matrix.
 We also explain the relevant  symmetries. Then we classify the projection matrices for
 the $\SO(1,2)$, $\SO(2)$ and $\SO(1,1)$ invariant equations.
   Sec.\ref{secBPS} contains our
 main results of  the BPS equations.
Sec.\ref{secBPSSO12} classifies the  $\SO(1,2)$ invariant  BPS equations preserving two,
four, six, eight, ten and twelve supersymmetries.\footnote{Note that in the present paper we focus on the sixteen ordinary supersymmetries
 and not the sixteen conformal supersymmetries.
For the BPS equations preserving conformal supersymmetries in super Yang-Mills
we refer the readers to Ref.\cite{Park:2006kt}.}  Sec.\ref{secBPSSO25} classifies the
  $\SO(2)$
invariant  BPS equations preserving two, four, six and  eight  supersymmetries.
In Sec.\ref{secBPSSO11} we
discuss the  $\SO(1,1)$ invariant  BPS equations which generalize the Nahm equations.
The final section, Sec.\ref{discussion} contains our results and discussions.
In  Appendix  we review the $\SO(8)$ triality and its relation to octonions.  \\

\textit{Note added}:
While this paper is being finished, Ref.\cite{KMlast} appears in
ArXiv which partially overlaps with our work, as it  discusses the BPS equations
of the form: $D_{y}X_{I}=\textstyle{\frac{1}{3!}}\cC_{IJKL}[X^{J},X^{K},X^{L}]$.
In the present paper, we  explicitly spell  the coefficients $\cC_{IJKL}$ and classify
various BPS equations.\\
{}\\

%%%%%%%%%%%%%%%%%%%%%%%%%%%%%%%%%%%%%%%%%%%%%%%%%%%%%%%%%%%%%%%%%%%%%%%%%%%%%%%%%%%%%%%%%%%%%%%%%%%%%%%%%%%%%
%%%%%%%%%%%%%%%%%%%%%%%%%%%%%%%%%%%%%%%%%%%%%%%%%%%%%%%%%%%%%%%%%%%%%%%%%%%%%%%%%%%%%%%%%%%%%%%%%%%%%%%%%%%%%
%%%%%%%%%%%%%%%%%%%%%%%%%%%%%%%%%%%%%%%%%%%%%%%%%%%%%%%%%%%%%%%%%%%%%%%%%%%%%%%%%%%%%%%%%%%%%%%%%%%%%%%%%%%%%
\section{Preliminaries\label{secPre}}
The multiple M2-brane theory has 8 real scalar fields $X^I, I=1,2,\cdots,8$
and a 16 component Majorana spinor
$\Psi$.
The supersymmetry transformation of the fermions in the Bagger-Lambert theory
assumes the form:
\be
\delta\Psi=\left(F_{\mu I}\Gamma^{\mu I}-\textstyle{\frac{1}{6}}F_{IJK}\Gamma^{IJK}\right)\varepsilon\,,
\label{BPS0}
\ee
where all the variables are three-algebra valued and we set
\be
\ba{ll}
F_{\mu I}\equiv D_{\mu} X_{I}\,,~~~~&~~~~~F_{IJK}\equiv\left[X_{I},X_{J},X_{K}\right]\,.
\ea
\ee
The bracket $[X_I,X_J,X_K]$ denotes the three-algebra product which is trilinear and totally
antisymmetric.
Note also that in contrast to the original convention \cite{BL} we let $I=1,2,\cdots,8$
and take $\mu\equiv 0,9,{10}$ directions as for the M2-brane  worldvolume for  convenience  to present the  BPS equations later,
\be
\ba{ccc}
x^{0}{\equiv t}\,,~~~~&~~~~x^{9}{\equiv x}\,,~~~~&~~~~x^{10}{\equiv y}\,.
\ea
\ee
The supersymmetry parameter is real and subject to the $\SO(1,2)$ projection condition:
\be
\Gamma^{txy}\varepsilon=\varepsilon\,,
\label{chiral12}
\ee
which is consistent with the opposite projection property, $\Gamma^{txy}\Psi=-\Psi$. Since the product of all the eleven-dimensional gamma matrices leads to the $32\times 32$ identity matrix $\Gamma^{txy123\cdots8}=1$, the above $\SO(1,2)$ projection condition coincides with the chirality condition of $\SO(8)$,
\be
\Gamma^{123\cdots 8}\varepsilon=\varepsilon\,.
\label{chiral8}
\ee
%%%%%%%%%%%%%%%%%%%%%%%%%%%%%%%%%%%%%%%%%%%%%%%%%%%%%%%%%%%%%%%%%%%%%%%%%%%%%%%%%%%%%%%%%%%%%%%%%%%%%%
%%%%%%%%%%%%%%%%%%%%%%%%%%%%%%%%%%%%%%%%%%%%%%%%%%%%%%%%%%%%%%%%%%%%%%%%%%%%%%%%%%%%%%%%%%%%%%%%%%%%%%
\subsection{Supersymmetry projection matrix - general}
In general for supersymmetric theories, the supersymmetry projection matrix $\Omega$ can be defined in terms of the commuting, real, orthonormal  supersymmetry parameters $\varepsilon_{1},\varepsilon_{2},\cdots,\varepsilon_{N}$,
\be
\ba{ll}
\displaystyle{\Omega:=\sum_{i=1}^{N} \varepsilon_{i}\varepsilon^{\dagger}_{i}\,,}~~~~&~~~~\varepsilon_{i}^{\dagger}\varepsilon_{j}=\delta_{ij}\,,
\ea
\ee
satisfying $\Omega^{\dagger}=\Omega^{2}=\Omega$. Here $N$ denotes  the number of the preserved supersymmetries,
\be
N=\Tr\Omega\,.
\ee
Naturally the eigenvalues of the projection matrices are either zero or one.\\

When the supersymmetry transformation of fermions  takes the form $\delta\Psi=\cF\varepsilon$ where $\cF$ denotes a bosonic quantity contracted with gamma matrices as in (\ref{BPS0}),
the general strategy to obtain the BPS equations is as follows~\cite{BPS68}:
\begin{enumerate}
\item \textit{Expand the projection matrix $\Omega$ in terms of the gamma matrix
product basis.}
\item \textit{Perform the matrix product $\cF\Omega$ and reexpress it in terms of
 the gamma matrix product basis.}
\item \textit{Read off the BPS equations from the coefficients of the linearly independent terms.}
\end{enumerate}
For example in  the Euclidean  four-dimensional minimal super Yang-Mills theory, we have two choices for the projection matrix $\Omega=\half(1\pm\gamma^{1234})$, while   $\cF=F_{ij}\gamma^{ij}$. Consequently, noting $\gamma^{12}\Omega=\mp\gamma^{34}\Omega$ \textit{etc.}, we get $F_{ij}\gamma^{ij}\Omega=2(F\mp{\star\,F})_{i4}\gamma^{i4}\Omega$ such that  the corresponding BPS equations are the well-known self-dual or anti-self-dual equations $F=\pm{\star\,F}$.  In this way, the complete classifications of the BPS equations in  six and eight-dimensional super Yang-Mills as well as the  pp-wave M-theory matrix model~\cite{Berenstein:2002jq} have been carried out \cite{BPS68,Park:2002cba,Kim:2002zg}. \\

 The present paper  concerns  the BPS equations of the Bagger-Lambert theory.
 Since the   eleven-dimensional spacetime  admits Majorana spinors we can set all
 the gamma matrices and the spinors to be real. In particular, the spatial gamma matrices are symmetric while the temporal gamma matrix is anti-symmetric.
 Consequently, also from (\ref{chiral8}),  the  projection matrices of the Bagger-Lambert theory  must satisfy
\be
\ba{lll}
\Omega=\Omega^{T}=\Omega^{\ast}\,,~~~~&~~~~\Omega=\Omega^{2}\,,~~~~&~~~~\Omega=\cP\Omega=\Omega\cP\,,
\ea
\label{Omegacon}
\ee
where $\cP$ is the $\SO(8)$ chiral projection matrix,
\be
\cP:=\half(1+\Gamma^{123\cdots 8})\,.
\label{defcP}
\ee
The most general form of such projection matrices reads
\be
\Omega=\left[c+ \Upsilon_{4}+\Gamma^{x}(c^{\prime}+\Upsilon_{4}^{\prime})
+\Gamma^{y}(c^{\prime\prime}+\Upsilon_{4}^{\prime\prime})+\Gamma^{xy}\Upsilon_{2}\right]\cP\,,
\label{GOME}
\ee
where $c,c^{\prime},c^{\prime\prime}$ are constants,
$\Upsilon_{4},\Upsilon_{4}^{\prime}, \Upsilon_{4}^{\prime\prime}$ are foursome productions of the $\SO(8)$ gamma matrices $\Gamma^{IJKL}$ contracted with  self-dual four-forms, and $\Upsilon_{2}$ is a twosome production of the  $\SO(8)$ gamma matrices $\Gamma^{IJ}$ contracted with a  two-form. All together, \textit{a priori}, there are ${3+3\times\half\left(\!\scriptsize{\ba{c}8\\4\ea}\!\right)+\left(\!\scriptsize{\ba{c}8\\2\ea}\!\right)=136}$ real  parameters which must be determined  by requiring the remaining condition $\Omega^{2}=\Omega$. The symmetry group  $\SO(1,2)\times\SO(8)$ in the Bagger-Lambert theory may reduce the number of the free parameters, but is not big enough to transform  all the free parameters, the two-form and the four-forms, into `canonical' forms. Note that the $\SO(8)$ rotation may take only one of $\left\{\Upsilon_{4},\Upsilon_{4}^{\prime}, \Upsilon_{4}^{\prime\prime},\Upsilon_{2}\right\}$ into a canonical form. In our choice, the canonical form of a two-form reads
\be
\Upsilon_{2}=a_{1}\Gamma^{12}+a_{2}\Gamma^{34}+a_{3}\Gamma^{56}+a_{4}\Gamma^{78}\,,
\ee
while the canonical form of a self-dual four-form reads
\be
\Upsilon_{4}=b_{1}\cE_{1}+b_{2}\cE_{2}+b_{3}\cE_{3}+b_{4}\cE_{4}+b_{5}\cE_{5}+b_{6}\cE_{6}+b_{7}\cE_{7}\,,
\label{canon4}
\ee
where we set
\begin{equation}
\begin{array}{cccc}
\cE_{1}=\Gamma_{8127}\cP\,,~&\cE_{2}=\Gamma_{8163}\cP\,,~&\cE_{3}=\Gamma_{8246}\cP\,,~& \cE_{4}=\Gamma_{8347}\cP\,,\\ {}&{}&{}&{}\\ \cE_{5}=\Gamma_{8567}\cP\,,~&\cE_{6}=\Gamma_{8253}\cP\,,~&
\cE_{7}=\Gamma_{8154}\cP\,.&{}
\end{array}
\label{Edef}
\end{equation}
The former is well known, while the latter is less familiar and we review it in Appendix~\ref{Appoct}. In (\ref{Edef}) the  subscript spatial indices  of the  gamma matrices are organized such that the three  indices after the common $8$ are  identical to those of
the totally anti-symmetric octonionic  structure constants~\cite{BPS68,Baez}:
\begin{equation}
\begin{array}{c}
~~e_{i}e_{j}=-\delta_{ij}+c_{ijk}\,e_{k}\,,~~~~~~~{i,j,k=1,2,\cdots,7\,,}\\ {}\\
1=c_{127}=c_{163}=c_{246}=c_{347}=c_{567}=c_{253}=c_{154}\,,~~~~~\mbox{others zero}\,.
\end{array}
\label{octconst}
\end{equation}
~\\

We say $\Omega$ is invariant under $\SO(2)$ rotation invariant on $xy$-plane if
$~[\Gamma_{\!xy}\,,\,\Omega]=0$. When this holds, for a finite angle $\phi$ and rotation
$G=e^{\phi\Gamma_{\!xy}}$,  from the equivalence
\be
\ba{lll}
\cF\Omega=0~~&~~~\Longleftrightarrow~~~&~~G\cF\Omega G^{-1}=G\cF G^{-1}\Omega=0\,,
\ea
\ee
we note that the corresponding BPS equations are, as a set, invariant under the rotation. Naturally this generalizes to an arbitrary subgroup of  $\SO(1,2){\times\SO(8)}$.\\

In the present paper instead of attempting to solve for the most general
 projection matrices,  we restrict  to the cases where $\Omega$ assumes the canonical form. Namely we focus on  two types of the BPS equations and classify the corresponding BPS equations completely: one is the $\SO(1,2)$ \textit{invariant cases}~\textit{i.e.}
\be
{\Omega=\left(c+ \Upsilon_{4}\right)\cP\,,}
\label{SO12Omega}
\ee
and  the other is the $\SO(2)^{\mathbf{5}}\equiv\SO(2){\times\SO(2)}{\times\SO(2)}{\times\SO(2)}{\times\SO(2)}$ \textit{invariant cases}~\textit{i.e.}
\be
{\Omega=\left(\mbox{constant}+\mbox{twosome~products~of~}\left\{\Gamma^{xy},\Gamma^{12},\Gamma^{34},
\Gamma^{56},\Gamma^{78}\right\}\right)\cP\,.}
\label{SO25Omega}
\ee
Here $\SO(1,2)$ and  $\SO(2)$  correspond to the M2 worldvolume Lorenz symmetry and the Cartan subgroup of the symmetry group  $\SO(1,2)\times\SO(8)$ respectively. In addition,  the former will easily   generate  various \textit{$\cM$-theoretic generalizations of the Nahm equations}  which are invariant under $\SO(1,1)\subset\SO(1,2)$, as the corresponding projection matrices are of the form:
\be
{\Omega=(1\pm\Gamma^{tx})\left(c+ \Upsilon_{4}\right)\cP\,.}
\label{SO11Omega}
\ee
~\\

%%%%%%%%%%%%%%%%%%%%%%%%%%%%%%%%%%%%%%%%%%%%%%%%%%%%%%%%%%%%%%%%%%%%%%%%%%%%%%%%%%%%%%%%%%%%%%%%%%%%%%
%%%%%%%%%%%%%%%%%%%%%%%%%%%%%%%%%%%%%%%%%%%%%%%%%%%%%%%%%%%%%%%%%%%%%%%%%%%%%%%%%%%%%%%%%%%%%%%%%%%%%%
\subsection{$\SO(1,2)$ invariant projection matrices}
The basic building blocks of all the possible $\SO(1,2)$ invariant  projection matrices are the following  $N=2$  projection matrices~\cite{BPS68}:
\begin{equation}
\Omega=\textstyle{\frac{1}{8}}\left(\cP+\salpha_{1}\salpha_{2}\cE_{1}+\salpha_{1}\salpha_{3}\cE_{2}
+\salpha_{3}\cE_{3}+\salpha_{2}\cE_{4}+\salpha_{1}\cE_{5}+\salpha_{1}\salpha_{2}\salpha_{3}\cE_{6}
+\salpha_{2}\salpha_{3}\cE_{7}\right)\,,
\label{SO12OG}
\end{equation}
where  $\alpha_{1},\,\alpha_{2},\,\alpha_{3}$ are  three independent signs,
\begin{equation}
\alpha_{1}^{2}=\alpha_{2}^{2}=\alpha_{3}^{2}=1\,.
\end{equation}
Three independent sign choices lead to eight possible combinations, hence eight  $N=2$ projection matrices.
They are orthogonal to each other and complete, as summing all of them gives an identity. Namely they form an orthogonal basis for the $\SO(1,2)$ invariant projection matrices. General $N=2k$ projection matrices  can be straightforwardly  obtained as a  $k$ sum of the above eight $N=2$ projection matrices.  Furthermore, from the $\SO(8)$
triality,  the ${8!}/[{k!(8{-k})!}]$ possibilities for the $k$ sum are all equivalent to each
other. The corresponding  $N{=2k}$ BPS equations are   $\SO(1,2){\times\SO(8{-k})}{\times\SO(k)}$ invariant.

%%%%%%
%%\be
%%F_{\mu I}=D_{\mu}X_{I}=0~~~~~~~~~\mbox{for~all~}\,\mu,~I\,.
%%\ee
%%%%%

%%%%%%%%%%%%%%%%%%%%%%%%%%%%%%%%%%%%%%%%%%%%%%%%%%%%%%%%%%%%%%%%%%%%%%%%%%%%%%%%%%%%%%%%%%%%%%%%%%%%%%
%%%%%%%%%%%%%%%%%%%%%%%%%%%%%%%%%%%%%%%%%%%%%%%%%%%%%%%%%%%%%%%%%%%%%%%%%%%%%%%%%%%%%%%%%%%%%%%%%%%%%%
\subsection{$\SO(2)$ invariant projection matrices}
The basic building blocks of all the possible $\SO(2)$ invariant projection
matrices are the following  $N=2$  projection matrices (see Appendix \ref{AppCartan}
for derivation):
\be
\ba{ll}
\Omega&=\textstyle{\frac{1}{8}}
\left[1+\Gamma^{xy}\!\left(\beta_{1}\Gamma^{12}+\beta_{2}\Gamma^{34}+\beta_{3}\Gamma^{56}+
\beta_{1}\beta_{2}\beta_{3}\Gamma^{78}\right)-\beta_{1}\beta_{2}\Gamma^{1234}
-\beta_{3}\beta_{1}\Gamma^{1256}-\beta_{2}\beta_{3}\Gamma^{1278}\right]\!\cP\\
{}&{}\\
{}&=\textstyle{\frac{1}{8}}(1+\beta_{1}\Gamma^{xy12})(1+\beta_{2}\Gamma^{xy34})(1+\beta_{3}\Gamma^{xy56})\cP\,,
\ea
\label{SO2OG}
\ee
where $\beta_{1}$, $\beta_{2}$, $\beta_{3}$ denote three independent signs,
\be
\beta_{1}^{2}=\beta_{2}^{2}=\beta_{3}^{2}=1\,.
\ee
Eight possible $N=2$ projection matrices form an orthogonal basis for
the $\SO(2)$  invariant projection matrices. General $N=2k$
projection matrices   can be straightforwardly  obtained as a  $k$ sum of
the above eight  $N=2$ projection matrices.  However, if the sum contains a pair
of  two  opposite overall sign factors~\textit{e.g.}  $\scriptstyle{(+++)}$ and
$\scriptstyle{(---)}$, the corresponding BPS configurations  become $\SO(1,2)$
invariant as $F_{\mu I}=0$ and  the BPS equations reduce to those  of $\SO(1,2)$
invariant BPS equations. Excluding these cases, up to $\SO(8)$ rotations,
there are five  inequivalent  $\SO(2)$ invariant projection matrices as follows.
\begin{itemize}
\item $N=2$ $\,\SO(2){\times\SU(4)}$ invariant projection matrix, with the choice of $\scriptstyle{(\beta_{1},\beta_{2},\beta_{3})\,=\,(+++)}$,
\be
\Omega=\textstyle{\frac{1}{8}}
\left[1+\Gamma^{xy}\!\left(\Gamma^{12}+\Gamma^{34}+\Gamma^{56}+\Gamma^{78}\right)-\Gamma^{1234}
-\Gamma^{1256}-\Gamma^{1278}\right]\cP\,.
\label{SO251}
\ee
\item $N=4$ $\,\SO(2){\times\SU(2)}{\times\SO(4)}$ invariant projection matrix, with  $\scriptstyle{(+++),(++-)}$,
\be
\Omega=\textstyle{\frac{1}{4}}
\left[1+\Gamma^{xy}\!\left(\Gamma^{12}+\Gamma^{34}\right)-\Gamma^{1234}\right]\cP\,.
\label{SO252}
\ee
\item $N=6$ $\,\SO(2){\times\SO(2)}{\times\SU(3)}$ invariant projection matrix, with  $\scriptstyle{(+++),(++-),(+-+)}$,
\be
\Omega=\textstyle{\frac{1}{8}}
\left[3+\Gamma^{xy}\!\left(3\Gamma^{12}+\Gamma^{34}+\Gamma^{56}-\Gamma^{78}\right)-\Gamma^{1234}
-\Gamma^{1256}+\Gamma^{1278}\right]\cP\,.
\label{SO253}
\ee
\item $N=8$ $\,\SO(2){\times\SO(2)}{\times\SO(6)}$ invariant projection matrix, with  $\scriptstyle{(+++),(++-),(+-+),(+--)}$,
\be
\Omega=\textstyle{\frac{1}{2}}(1+\Gamma^{xy12})\cP\,.
\label{SO254}
\ee
\item $N=8$ $\,\SO(2){\times\SU(4)}$ invariant projection matrix,  with $\scriptstyle{(+++),(++-),(+-+),(-++)}$,
\be
\Omega=\textstyle{\frac{1}{4}}
\left[2+\Gamma^{xy}\!\left(\Gamma^{12}+\Gamma^{34}+\Gamma^{56}-\Gamma^{78}\right)\right]\cP\,.
\label{SO255}
\ee

\end{itemize}

%%%%%%%%%%%%%%%%%%%%%%%%%%%%%%%%%%%%%%%%%%%%%%%%%%%%%%%%%%%%%%%%%%%%%%%%%%%%%%%%%%%%%%%%%%%%%%%%%%%%%%
%%%%%%%%%%%%%%%%%%%%%%%%%%%%%%%%%%%%%%%%%%%%%%%%%%%%%%%%%%%%%%%%%%%%%%%%%%%%%%%%%%%%%%%%%%%%%%%%%%%%%%
\subsection{$\SO(1,1)$  invariant projection matrices}
For $\SO(1,1)$ invariant  projection matrices, we have the following $N=1$ projection matrices:
\begin{equation}
\Omega=\textstyle{\frac{1}{16}}\left(1+\alpha_{0}\Gamma^{tx}\right)
\left(\cP+\salpha_{1}\salpha_{2}\cE_{1}+\salpha_{1}\salpha_{3}\cE_{2}
+\salpha_{3}\cE_{3}+\salpha_{2}\cE_{4}+\salpha_{1}\cE_{5}+\salpha_{1}\salpha_{2}\salpha_{3}\cE_{6}
+\salpha_{2}\salpha_{3}\cE_{7}\right)\,,
\label{SO11OG}
\end{equation}
where  $\alpha_{0},\,\alpha_{1},\,\alpha_{2},\,\alpha_{3}$ are  four independent signs,
\begin{equation}
\alpha_{0}^2=\alpha_{1}^{2}=\alpha_{2}^{2}=\alpha_{3}^{2}=1\,.
\end{equation}
Sixteen possible $N=1$ projection matrices form  an  orthogonal basis for the $\SO(1,1)$  invariant projection matrices. Generic $N=k$ $\SO(1,1)$  invariant projection matrices  may be   obtained  straightforwardly as a  $k$ sum of the above sixteen $N=1$ projection matrices.   For each sum, we may decompose
\be
\ba{lll}
N=N_{+}+N_{-}\,,~~~~&~~~~N_{+}=n_{+}+n\,,~~~~&~~~~N_{-}=n_{-}+n\,,
\ea
\label{Npmdef}
\ee
such that $N_{\pm}$  denotes the number of $N=1$ projection matrices in the sum whose  $\alpha_{0}$ values are $\pm 1$, and $n$ counts the number of $N=1$ projection matrix pairs which have the same $\alpha_{1},\,\alpha_{2},\,\alpha_{3}$ values and opposite $\alpha_{0}$ signs. There are   ${8!}/[{n_{+}!n_{-}!n!(8{-n_{+}}{-n_{-}}{-n})!}]$ possibilities for the sum which are all equivalent to another, thanks to  the $\SO(8)$ triality.
Furthermore, if $n$ is nontrivial~$n\neq 0$, then the BPS configurations become $\SO(1,2)$ invariant as $F_{\mu I}=0$ and the number of  the preserved supersymmetries is automatically increased from $n_{+}+n_{-}+2n$ to $2(n_{+}+n_{-}+n)$. In this case the BPS equations reduce to those  of $\SO(1,2)$ invariant BPS equations. Genuinely $\SO(1,1)$ invariant  BPS equations appear only when $n=0$. The corresponding  $(N_{+},N_{-})$ BPS equations are then $\SO(1,1){\times\SO(N_{+})}{\times\SO(N_{-})}{\times\SO(8{-N_{+}}{-N_{-}})}$ invariant with the natural  restriction  $N_{+}{+N_{-}}\leq 8$.

%%%%%
%%\begin{equation}
%%\Omega=\half(1+\salpha_{0}\sigma_{3})\otimes
%%\textstyle{\frac{1}{8}}\left(P+\salpha_{1}\salpha_{2}E_{1}+\salpha_{1}\salpha_{3}E_{2}
%%+\salpha_{3}E_{3}+\salpha_{2}E_{4}+\salpha_{1}E_{5}+\salpha_{1}\salpha_{2}\salpha_{3}E_{6}
%%+\salpha_{2}\salpha_{3}E_{7}\right)\,, \label{Omega}
%%\end{equation}
%%where  $\alpha_{0},\,\alpha_{1},\,\alpha_{2},\,\alpha_{3}$ are  four independent signs,
%%\begin{equation}
%%1=\alpha_{0}^{2}=\alpha_{1}^{2}=\alpha_{2}^{2}=\alpha_{3}^{2}\,.
%%\end{equation}
%%%%%
%%%%%%%%%%%%%%%%%%%%%%%%%%%%%%%%%%%%%%%%%%%%%%%%%%%%%%%%%%%%%%%%%%%%%%%%%%%%%%%%%%%%%%%%%%%%%%%%
%%%%%%%%%%%%%%%%%%%%%%%%%%%%%%%%%%%%%%%%%%%%%%%%%%%%%%%%%%%%%%%%%%%%%%%%%%%%%%%%%%%%%%%%%%%%%%%%
\section{Classification of the BPS equations\label{secBPS}}
%%%%%%%%%%%%%%%%%%%%%%%%%%%%%%%%%%%%%%%%%%%%%%%%%%%%%%%%%%%%%%%%%%%%%%%%%%%%%%%%%%%%%%%%%%%%%%%%
%%%%%%%%%%%%%%%%%%%%%%%%%%%%%%%%%%%%%%%%%%%%%%%%%%%%%%%%%%%%%%%%%%%%%%%%%%%%%%%%%%%%%%%%%%%%%%%%
\subsection{$\SO(1,2)$ invariant BPS equations\label{secBPSSO12}}
The generic $N=2$ $\SO(1,2)$ invariant projection matrix~(\ref{SO12OG}) leads to the following
 $N{=2}$ $\,\SO(1,2){\times\SO(7)}$ invariant BPS equations which involve   three free sign factors  $\,\alpha_{1}^{2}=\alpha_{2}^{2}=\alpha_{3}^{2}=1$:
\be
\ba{c}
F_{\mu I}=0\,,~~~~~~~~~~~\mu=t,x,y\,,~~~~~I=1,2,\cdots,8\,,
\label{FmuIzero}
\ea
\ee
and
\be
\ba{l}
{\alpha_{1}\alpha_{2}}F_{278}+{\alpha_{2}\alpha_{3}}F_{548}+{\alpha_{3}\alpha_{1}}F_{638}
+\alpha_{1}F_{234}+\alpha_{2}F_{256}+\alpha_{3}F_{357}+{\alpha_{1}\alpha_{2}\alpha_{3}}F_{476}=0\,,\\
{}\\
{\alpha_{1}\alpha_{2}}F_{718}+{\alpha_{2}\alpha_{3}}F_{376}+{\alpha_{3}\alpha_{1}}F_{475}
+\alpha_{1}F_{143}+\alpha_{2}F_{165}+\alpha_{3}F_{468}+{\alpha_{1}\alpha_{2}\alpha_{3}}F_{538}=0\,,\\
{}\\
{\alpha_{1}\alpha_{2}}F_{456}+{\alpha_{2}\alpha_{3}}F_{267}+{\alpha_{3}\alpha_{1}}F_{168}
+\alpha_{1}F_{124}+\alpha_{2}F_{478}+\alpha_{3}F_{517}+{\alpha_{1}\alpha_{2}\alpha_{3}}F_{258}=0\,,\\
{}\\
{\alpha_{1}\alpha_{2}}F_{536}+{\alpha_{2}\alpha_{3}}F_{158}+{\alpha_{3}\alpha_{1}}F_{257}
+\alpha_{1}F_{132}+\alpha_{2}F_{738}+\alpha_{3}F_{628}+{\alpha_{1}\alpha_{2}\alpha_{3}}F_{167}=0\,,\\
{}\\
{\alpha_{1}\alpha_{2}}F_{346}+{\alpha_{2}\alpha_{3}}F_{418}+{\alpha_{3}\alpha_{1}}F_{427}
+\alpha_{1}F_{678}+\alpha_{2}F_{126}+\alpha_{3}F_{137}+{\alpha_{1}\alpha_{2}\alpha_{3}}F_{328}=0\,,\\
{}\\
{\alpha_{1}\alpha_{2}}F_{354}+{\alpha_{2}\alpha_{3}}F_{273}+{\alpha_{3}\alpha_{1}}F_{318}
+\alpha_{1}F_{758}+\alpha_{2}F_{152}+\alpha_{3}F_{248}+{\alpha_{1}\alpha_{2}\alpha_{3}}F_{174}=0\,,\\
{}\\
{\alpha_{1}\alpha_{2}}F_{128}+{\alpha_{2}\alpha_{3}}F_{236}+{\alpha_{3}\alpha_{1}}F_{245}
+\alpha_{1}F_{568}+\alpha_{2}F_{348}+\alpha_{3}F_{153}+{\alpha_{1}\alpha_{2}\alpha_{3}}F_{146}=0\,,\\
{}\\
{\alpha_{1}\alpha_{2}}F_{127}+{\alpha_{2}\alpha_{3}}F_{154}+{\alpha_{3}\alpha_{1}}F_{163}
+\alpha_{1}F_{567}+\alpha_{2}F_{347}+\alpha_{3}F_{246}+{\alpha_{1}\alpha_{2}\alpha_{3}}F_{253}=0\,.
\ea
\label{BPSMasterSO12}
\ee
In particular, the $\SO(1,2)$ invariance, the M2-brane worldvolume Lorentz symmetry, removes any  worldvolume dependence, $D_{\mu}X_{I}=0$ for all $\mu$ and $I$. \\

The above set of  BPS equations can be regarded as the master equations since any $N{=2k}$ BPS equations can be obtained by imposing $k$ copies of distinct $(\alpha_{1},\alpha_{2},\alpha_{3})$ choices.    The corresponding  $N{=2k}$ BPS equations are then $\SO(1,2){\times\SO(8{-k})}{\times\SO(k)}$ invariant. We find for $N{=14}$ and $N{=16}$ the corresponding BPS equations are trivial, $\,F_{\mu I}=F_{IJK}=0$. Other nontrivial cases are as follows.

%%%%%%%%%%%%%%%%%%%%%%%%%%%%%%%%%%%%%%%%%%%%%%%%%%%%%%%%%%%%%%%%%%%%%%%%%%%%%%%%%%%%%%%%%%%%%%%%%%%%%%%%%%%%%%%%
%%%%%%%%%%%%%%%%%%%%%%%%%%%%%%%%%%%%%%%%%%%%%%%%%%%%%%%%%%%%%%%%%%%%%%%%%%%%%%%%%%%%%%%%%%%%%%%%%%%%%%%%%%%%%%%%
\subsubsection{$N=2$  $\,\SO(1,2){\times\SO(7)}$ invariant  BPS equations - \textit{octonion}}
With the choice of ${(\alpha_{1},\alpha_{2},\alpha_{3})=(+++)}$,
the $N{=2}$  $\,\SO(1,2){\times\SO(7)}$ invariant  BPS equations~(\ref{FmuIzero}), (\ref{BPSMasterSO12}) assume  a compact form:
\be
\ba{ll}
F_{\mu I}=0\,,~~~~~&~~~~\cC_{IJKL}F^{JKL}=0\,,
\ea
\label{SO12N2}
\ee
where $\cC_{IJKL}$ is a $\SO(7)$ invariant  four-form in eight dimensions, defined  in terms of the octonionic structure constant (\ref{octconst}),
\be
\ba{lll}
\cC_{ijk8}\equiv c_{ijk}\,,~~~~&~~~~\cC_{ijkl}\equiv\textstyle{\frac{1}{6}}\epsilon_{pqrijkl}c_{pqr}~~~~~~~&\mbox{where}~~~~~~
1\leq i,j,k,l\leq 7\,.
\ea
\ee
~\\

BPS states preserving  $N{=2k}$ supersymmetries  then satisfy $k$ copies of the $N{=2}$ BPS equations of different $\alpha$ choices.  The corresponding  $N{=2k}$ BPS equations are $\SO(1,2){\times\SO(k)}{\times\SO(8{-k})}$ invariant, and involve  $k$ different octonionic structures.

%%%%%%%%%%%%%%%%%%%%%%%%%%%%%%%%%%%%%%%%%%%%%%%%%%%%%%%%%%%%%%%%%%%%%%%%%%%%%%%%%%%%%%%%%%%%%%%%%%%%%%%%%%%%%%%%
%%%%%%%%%%%%%%%%%%%%%%%%%%%%%%%%%%%%%%%%%%%%%%%%%%%%%%%%%%%%%%%%%%%%%%%%%%%%%%%%%%%%%%%%%%%%%%%%%%%%%%%%%%%%%%%%
\subsubsection{$N=4$  $\,\SO(1,2){\times\SO(6)}{\times\SO(2)}$ invariant  BPS equations - \textit{complex}}
The $N{=4}$  $\,\SO(1,2){\times\SO(6)}{\times\SO(2)}$ invariant  BPS equations are, with $F_{\mu I}{=0}$,
\be
\ba{ll}
F_{IJK}\cJ^{JK}=0\,,~~~&~~~F_{IJK}=(1{\otimes\cJ}{\otimes\cJ}+\cJ{\otimes 1}{\otimes\cJ}+
\cJ{\otimes\cJ}{\otimes 1})_{IJK}{}^{LMN}F_{LMN}\,,
\ea
\label{SO12N4}
\ee
where $\cJ$ is a complex structure $\cJ^{2}=-1$, $\cJ^{T}=-\cJ$ and hence $\SU(4){\times\SO(2)}$ invariant.\\

With the specific choice of $\alpha$'s as $\scriptstyle{(+++),(++-)}$, one gets
\be
\half\cJ_{IJ}\Gamma^{IJ}=\Gamma^{12}+\Gamma^{34}+\Gamma^{56}+\Gamma^{78}\,.
\label{cJrep}
\ee
{}\\

In terms of the corresponding   holomorphic, anti-holomorphic coordinates $a,\bar{a}=1,2,3,4$ and the metric $\delta^{a\bar{a}}$,   the above
$N{=4}$  $\,{\SO(1,2)}{\times\SO(6)}{\times\SO(2)}$   BPS equations~(\ref{SO12N4}) can be rewritten as
\be
\ba{ll}
F_{ab}{}^{b}=F_{\bar{a}b}{}^{b}=0\,,~~~~~&~~~~F_{abc}=F_{\bar{a}\bar{b}\bar{c}}=0\,.
\ea
\ee
Namely  $F_{(1,2)}$, $F_{(2,1)}$ are primitive and  $F_{(3,0)}{=F_{(0,3)}}{=0}$. \\

We note that summing two  $N{=2}$ projection matrices generates one complex structure. Hence in general,
summing $k>2$ of  $N{=2}$ projection matrices will present \mbox{$\left(\!\scriptsize{\ba{c}k\\2\ea}\!\right)$} number of complex structures to the corresponding $\SO(1,2){\times\SO(8{-k})}{\times\SO(k)}$ invariant  BPS equations. The $\half k(k-1)$  complex structures form singlets under $\SO(8{-k})$ and are in the adjoint representation or $k$-dimensional two-form representation   of $\SO(k)$. In fact, they correspond to the generators of $\SO(k)$. Nevertheless, the corresponding   $\half k(k-1)$  number of complex structures are degenerate in the sense that  distinct  $[\frac{k+1}{2}]$ of them  are sufficient to lead to  the full $N{=2k}$ BPS equations.

%%%%%%%%%%%%%%%%%%%%%%%%%%%%%%%%%%%%%%%%%%%%%%%%%%%%%%%%%%%%%%%%%%%%%%%%%%%%%%%%%%%%%%%%%%%%%%%%%%%%%%%%%%%%%%%%
%%%%%%%%%%%%%%%%%%%%%%%%%%%%%%%%%%%%%%%%%%%%%%%%%%%%%%%%%%%%%%%%%%%%%%%%%%%%%%%%%%%%%%%%%%%%%%%%%%%%%%%%%%%%%%%%
\subsubsection{$N=6$  $\,{\SO(1,2)}{\times\SO(5)}{\times\SO(3)}$ invariant  BPS equations - \textit{quarternion}}
The $N{=6}$  $\,{\SO(1,2)}{\times\SO(5)}{\times\SO(3)}$ invariant  BPS equations are, with $F_{\mu I}{=0}$,
\be
F_{IJK}\cJ_{p}^{JK}=0\,,~~~~~~~~p=1,2,3\,,
\label{SO12N6}
\ee
where $\cJ_{1},\cJ_{2},\cJ_{3}$ are three distinct complex structures satisfying the quaternion relations:
\be
\cJ_{1}^{2}=\cJ_{2}^{2}=\cJ_{3}^{2}=\cJ_{1}\cJ_{2}\cJ_{3}=-1\,.
\ee
It is worth to note that the remaining relation  of (\ref{SO12N4}) \textit{i.e.} $F_{(3,0)}{=0}$  is fulfilled   automatically for each complex structure. \\

With the specific choice of $\alpha$'s as $\scriptstyle{(+++),(++-),(+-+)}$, one gets
\be
\ba{l}
\half\cJ_{1}^{IJ}\Gamma_{IJ}=\Gamma^{12}+\Gamma^{34}+\Gamma^{56}+\Gamma^{78}\,,\\
{}\\
\half\cJ_{2}^{IJ}\Gamma_{IJ}=\Gamma^{14}+\Gamma^{23}+\Gamma^{58}+\Gamma^{67}\,,\\
{}\\
\half\cJ_{3}^{IJ}\Gamma_{IJ}=\Gamma^{13}+\Gamma^{42}+\Gamma^{57}+\Gamma^{86}\,.
\ea
\label{J3}
\ee
~\\

Summing three  $N{=2}$ projection matrices generates one quarternion  structure. Hence in general,
summing ${k>3}$ of  $N{=2}$ projection matrices will present \mbox{$\left(\!\scriptsize{\ba{c}k\\3\ea}\!\right)$} number of quarternion  structures to the corresponding $\SO(1,2){\times\SO(8{-k})}{\times\SO(k)}$ invariant  BPS equations. The \mbox{$\left(\!\scriptsize{\ba{c}k\\3\ea}\!\right)$} quarternion  structures are singlets under $\SO(8{-k})$ and form a  $k$-dimensional three-form representation of $\SO(k)$.  Nevertheless, the corresponding   $\textstyle{\frac{1}{6}}k(k-1)(k-2)$  number of quarternion  structures are degenerate in the sense that  distinct   $[\frac{k+2}{3}]$ of them  are sufficient   to give  the full $N{=2k}$ BPS equations.

%%%%%%%%%%%%%%%%%%%%%%%%%%%%%%%%%%%%%%%%%%%%%%%%%%%%%%%%%%%%%%%%%%%%%%%%%%%%%%%%%%%%%%%%%%%%%%%%%%%%%%%%%%%%%%%%
%%%%%%%%%%%%%%%%%%%%%%%%%%%%%%%%%%%%%%%%%%%%%%%%%%%%%%%%%%%%%%%%%%%%%%%%%%%%%%%%%%%%%%%%%%%%%%%%%%%%%%%%%%%%%%%%
\subsubsection{$N=8$  $\,{\SO(1,2)}{\times\SO(4)}{\times\SO(4)}$ invariant  BPS equations}
The $N{=8}$  $\,{\SO(1,2)}{\times\SO(4)}{\times\SO(4)}$ invariant  BPS equations are, with $F_{\mu I}{=0}$,
\be
F_{IJK}+\half F_{I}{}^{LM}\cT_{JKLM}+\half F_{J}{}^{LM}\cT_{KILM}+\half F_{K}{}^{LM}\cT_{IJLM}=0\,,
\label{SO12N8}
\ee
%%%%%
%%\be
%%F_{IJK}+\half\cT_{IJLM}F_{K}{}^{LM}+\half\cT_{JKLM}F_{I}{}^{LM}+\half\cT_{KILM}F_{J}{}^{LM}=0\,,
%%\label{SO12N8}
%%\ee
%%%%%
where $\cT_{IJKL}$ is a $\SO(4)\times\SO(4)$ invariant self-dual four-form. With the specific choice of $\alpha$'s as $\scriptstyle{(+++),(++-),(+-+),(+--)}$, one gets
\be
\textstyle{\frac{1}{4!}}\cT_{IJKL}\Gamma^{IJKL}=\Gamma^{1234}+\Gamma^{5678}\,.
\label{Tsdff}
\ee
~\\

Summing four  $N{=2}$ projection matrices generates one self-dual four-form  structure. Hence in general,
summing $k>4$ of  $N{=2}$ projection matrices will present \mbox{$\left(\!\scriptsize{\ba{c}k\\4\ea}\!\right)$} number of self-dual four-form   structures to the corresponding $\SO(1,2){\times\SO(8{-k})}{\times\SO(k)}$ invariant  BPS equations. The \mbox{$\left(\!\scriptsize{\ba{c}k\\4\ea}\!\right)$}  self-dual four-form  structures are singlets under $\SO(8{-k})$ and form a  $k$-dimensional four-form representation of $\SO(k)$.  Nevertheless, the corresponding   $\textstyle{\frac{k!}{4!(k-4)!}}$  number of self-dual four-forms   are degenerate in the sense that  distinct   $[\frac{k+3}{4}]$ of them  are sufficient   to give  the full $N{=2k}$ BPS equations.

%%%%%%%%%%%%%%%%%%%%%%%%%%%%%%%%%%%%%%%%%%%%%%%%%%%%%%%%%%%%%%%%%%%%%%%%%%%%%%%%%%%%%%%%%%%%%%%%%%%%%%%%%%%%%%%%
%%%%%%%%%%%%%%%%%%%%%%%%%%%%%%%%%%%%%%%%%%%%%%%%%%%%%%%%%%%%%%%%%%%%%%%%%%%%%%%%%%%%%%%%%%%%%%%%%%%%%%%%%%%%%%%%
\subsubsection{$N=10$  $\,{\SO(1,2)}{\times\SO(3)}{\times\SO(5)}$ invariant  BPS equations}
For  $N{=10}$  $\,{\SO(1,2)}{\times\SO(3)}{\times\SO(5)}$ case there seems no novel structure to appear.  One economic fashion to write the $N=10$  $\,{\SO(1,2)}{\times\SO(3)}{\times\SO(5)}$ invariant  BPS equations is to employ  a $\SO(4)\times\SO(4)$ invariant  self-dual four-form and a complex structure:  with $F_{\mu I}{=0}$,\footnote{
Alternatively we can express them in terms of two sets of \textit{either} $\SO(4)\times\SO(4)$ invariant  self-dual four-forms one given by (\ref{Tsdff}) the other by $
\half({\Gamma_{1234}}{+\Gamma_{5678}}{+\Gamma_{1256}}{+\Gamma_{3478}}{+\Gamma_{1357}}{+\Gamma_{2468}}
{+\Gamma_{1467}}{+\Gamma_{2358}})$ \textit{\,or\,} quarternionic complex structures one by (\ref{J3}) and the other by
$\Gamma_{14}{+\Gamma_{85}}{+\Gamma_{76}}{+\Gamma_{23}}$, $\,\Gamma_{15}{+\Gamma_{48}}{+\Gamma_{73}}{+\Gamma_{62}}$,
$\,\Gamma_{18}{+\Gamma_{54}}{+\Gamma_{72}}{+\Gamma_{36}}$.}
\be
\ba{ll}
F_{IJK}+\textstyle{\frac{3}{2}}F_{[I}{}^{LM}\cT_{JK]LM}=0\,,~~~~~&~~~~~
F_{IJK}\cJ^{JK}=0\,.
\label{SO12N10}
\ea
\ee
The specific choice of $\alpha$'s as $\scriptstyle{(+++),(++-),(+-+),(+--),(-++)}$ gives
\be
\ba{ll}
\textstyle{\frac{1}{4!}}\cT_{IJKL}\Gamma^{IJKL}=\Gamma^{1234}+\Gamma^{5678}\,,~~~~&~~~~
\half\cJ_{IJ}\Gamma^{IJ}=\Gamma^{18}-\Gamma^{27}+\Gamma^{36}-\Gamma^{45}\,.
\ea
\label{TJ}
\ee

%%%%%%%%%%%%%%%%%%%%%%%%%%%%%%%%%%%%%%%%%%%%%%%%%%%%%%%%%%%%%%%%%%%%%%%%%%%%%%%%%%%%%%%%%%%%%%%%%%%%%%%%%%%%%%%%
%%%%%%%%%%%%%%%%%%%%%%%%%%%%%%%%%%%%%%%%%%%%%%%%%%%%%%%%%%%%%%%%%%%%%%%%%%%%%%%%%%%%%%%%%%%%%%%%%%%%%%%%%%%%%%%%
\subsubsection{$N=12$  $\,{\SO(1,2)}{\times\SO(2)}{\times\SO(6)}$ invariant  BPS equations}
The $N{=12}$  $\,{\SO(1,2)}{\times\SO(2)}{\times\SO(6)}$ invariant  BPS equations are,
with $F_{\mu I}{=0}$,\footnote{Of course, the above $N=12$ BPS equations can be obtained by imposing a pair of two distinct  quarternionic BPS equations~(\ref{SO12N6}). There are  $\half\!\left(\!\!\scriptsize{\ba{c}6\\3\ea}\!\!\right)=10$ such pairs and any of them leads to the same $N=12$ BPS equations. For example we may choose one quarternion  structure   from (\ref{J3}) and the other  by
$\Gamma^{12}{+\Gamma^{87}}{+\Gamma^{56}}{+\Gamma^{43}}$,
$~\Gamma^{17}{+\Gamma^{28}}{+\Gamma^{53}}{+\Gamma^{64}}$,
$~\Gamma^{18}{+\Gamma^{72}}{+\Gamma^{54}}{+\Gamma^{36}}$, corresponding to the $\alpha$ choices $\scriptstyle{(+++),(++-),(+-+)}$ and  $\scriptstyle{(+--),(-++),(-+-)}$.}
\be
F_{IJK}\cT_{p}^{JK}=0\,,~~~~~~~~p=1,2,3,4,5,6\,,
\label{SO12N12}
\ee
where $\cT_{p}^{IJ}$'s are  $\SO(2)\times\SO(6)$ covariant two-forms: fundamental under $\SO(6)$ and singlet under $\SO(2)$. With the specific choice of $\alpha$'s as $\scriptstyle{(+++),(++-),(+-+),(+--),(-++),(-+-)}$, one gets
\be
\ba{ll}
\textstyle{\frac{1}{2}}\cT_{1}^{IJ}\Gamma_{IJ}=\Gamma^{14}+\Gamma^{23}\,,~~~~~&~~~~~
\textstyle{\frac{1}{2}}\cT_{2}^{IJ}\Gamma_{IJ}=\Gamma^{67}+\Gamma^{58}\,,\\
{}&{}\\
\textstyle{\frac{1}{2}}\cT_{3}^{IJ}\Gamma_{IJ}=\Gamma^{16}+\Gamma^{25}\,,~~~~~&~~~~~
\textstyle{\frac{1}{2}}\cT_{4}^{IJ}\Gamma_{IJ}=\Gamma^{74}+\Gamma^{83}\,,\\
{}&{}\\
\textstyle{\frac{1}{2}}\cT_{5}^{IJ}\Gamma_{IJ}=\Gamma^{17}+\Gamma^{28}\,,~~~~~&~~~~~
\textstyle{\frac{1}{2}}\cT_{6}^{IJ}\Gamma_{IJ}=\Gamma^{35}+\Gamma^{46}\,.
\ea
\label{T6def}
\ee
~\\

%%%%%%%%%%%%%%%%%%%%%%%%%%%%%%%%%%%%%%%%%%%%%%%%%%%%%%%%%%%%%%%%%%%%%%%%%%%%%%%%%%%%%%%%%%%%%%%%
%%%%%%%%%%%%%%%%%%%%%%%%%%%%%%%%%%%%%%%%%%%%%%%%%%%%%%%%%%%%%%%%%%%%%%%%%%%%%%%%%%%%%%%%%%%%%%%%
\subsection{$\SO(2)$ invariant BPS equations\label{secBPSSO25}}
The generic $N=2$  projection matrix~(\ref{SO2OG}) leads to the following
 $N{=2}$ $\,\SO(2){\times\SU(4)}$ invariant BPS equations which involve     three free sign factors $\beta_{1}^{2}=\beta_{2}^{2}=\beta_{3}^{2}=1$:
 \be
\ba{llll}
F_{x1}+\beta_{1}F_{y2}=0\,,~~~&~~~~F_{x3}+\beta_{2}F_{y4}=0\,,~~~&~~~~
F_{x5}+\beta_{3}F_{y6}=0\,,~~~&~~~~F_{x7}+\beta_{1}\beta_{2}\beta_{3}F_{y8}=0\,,\\
{}&{}&{}&{}\\
F_{x2}-\beta_{1}F_{y1}=0\,,~~~&~~~~F_{x4}-\beta_{2}F_{y3}=0\,,~~~&~~~~
F_{x6}-\beta_{3}F_{y5}=0\,,~~~&~~~~F_{x8}-\beta_{1}\beta_{2}\beta_{3}F_{y7}=0\,,
\ea
\label{SO25BPSeqMaster1}
\ee
and
\be
\ba{ll}
{F_{t1}+\beta_{2}F_{134}+\beta_{3}F_{156}+\beta_{1}\beta_{2}\beta_{3}F_{178}=0\,,}~~~&~~~
{F_{135}-\beta_{1}\beta_{2}F_{245}-\beta_{2}\beta_{3}F_{146}-\beta_{3}\beta_{1}F_{236}=0\,,}\\
{}&{}\\
{F_{t2}+\beta_{2}F_{234}+\beta_{3}F_{256}+\beta_{1}\beta_{2}\beta_{3}F_{278}=0\,,}~~~&~~~
{F_{136}-\beta_{1}\beta_{2}F_{246}+\beta_{2}\beta_{3}F_{145}+\beta_{3}\beta_{1}F_{235}=0\,,}\\
{}&{}\\
{F_{t3}+\beta_{1}F_{312}+\beta_{3}F_{356}+\beta_{1}\beta_{2}\beta_{3}F_{378}=0\,,}~~&~~~
{F_{137}-\beta_{1}\beta_{2}F_{247}-\beta_{2}\beta_{3}F_{238}-\beta_{3}\beta_{1}F_{148}=0\,,}\\
{}&{}\\
{F_{t4}+\beta_{1}F_{412}+\beta_{3}F_{456}+\beta_{1}\beta_{2}\beta_{3}F_{478}=0\,,}~~&~~~
{F_{138}-\beta_{1}\beta_{2}F_{248}+\beta_{2}\beta_{3}F_{237}+\beta_{3}\beta_{1}F_{147}=0\,,}\\
{}&{}\\
{F_{t5}+\beta_{1}F_{512}+\beta_{2}F_{534}+\beta_{1}\beta_{2}\beta_{3}F_{578}=0\,,}~~&~~~
{F_{157}-\beta_{1}\beta_{2}F_{168}-\beta_{2}\beta_{3}F_{258}-\beta_{3}\beta_{1}F_{267}=0\,,}\\
{}&{}\\
{F_{t6}+\beta_{1}F_{612}+\beta_{2}F_{634}+\beta_{1}\beta_{2}\beta_{3}F_{678}=0\,,}~~&~~~
{F_{158}+\beta_{1}\beta_{2}F_{167}+\beta_{2}\beta_{3}F_{257}-\beta_{3}\beta_{1}F_{268}=0\,,}\\
{}&{}\\
{F_{t7}+\beta_{1}F_{712}+\beta_{2}F_{734}+\beta_{3}F_{756}=0\,,}~~&~~~
{F_{357}-\beta_{1}\beta_{2}F_{368}-\beta_{2}\beta_{3}F_{467}-\beta_{3}\beta_{1}F_{458}=0\,,}\\
{}&{}\\
{F_{t8}+\beta_{1}F_{812}+\beta_{2}F_{834}+\beta_{3}F_{856}=0\,,}~~&~~~
{F_{358}+\beta_{1}\beta_{2}F_{367}-\beta_{2}\beta_{3}F_{468}+\beta_{3}\beta_{1}F_{457}=0\,.}
\ea
\label{SO25BPSeqMaster2}
\ee
The above set of  BPS equations can be regarded as the master equations since any $N{=2k}$ $\SO(2)^{\mathbf{5}}$ invariant BPS equations corresponding to the projection matrices (\ref{SO251}\,-\,\ref{SO255}) can be obtained by imposing $k$ copies of distinct $(\beta_{1},\beta_{2},\beta_{3})$ choices.  We find,
among them,  the  $N=8$ $\,\SO(2){\times\SU(4)}$ invariant projection matrix~(\ref{SO255}) leads   to the trivial BPS configuration $F_{\mu I}=F_{IJK}=0$.  Other nontrivial cases are as follows.
%%%%%%%%%%%%%%%%%%%%%%%%%%%%%%%%%%%%%%%%%%%%%%%%%%%%%%%%%%%%%%%%%%%%%%%%%%%%%%%%%%%%%%%%%%%%%%%%%%%
%%%%%%%%%%%%%%%%%%%%%%%%%%%%%%%%%%%%%%%%%%%%%%%%%%%%%%%%%%%%%%%%%%%%%%%%%%%%%%%%%%%%%%%%%%%%%%%%%%%
 \subsubsection{$N=2$ $\,\SO(2){\times\SU(4)}$ invariant BPS equations}
The $N=2$ $\,\SO(2){\times\SU(4)}$ invariant  BPS equations corresponding to the projection matrix (\ref{SO251}) or  the choice  ${(\beta_{1},\beta_{2},\beta_{3})=(+++)}$ in (\ref{SO25BPSeqMaster1}) and (\ref{SO25BPSeqMaster2}) assume a compact form, up to Hermitian conjugation:
\be
\ba{lll}
F_{z\bar{a}}=0\,,~~~~~&~~~~~F_{ta}-iF_{ab}{}^{b}=0\,,~~~~~&~~~~~F_{abc}=0\,,
\ea
\ee
 provided we complexify the $\SO(8)$ coordinates  by  the complex structure ${\Gamma_{12}}{+\Gamma_{34}}{+\Gamma_{56}}{+\Gamma_{78}}$,  to introduce the holomorphic and anti-holomorphic variables  $a,\bar{a}=1,2,3,4$ such that the metric is $\delta_{a\bar{a}}$ and
\be
\ba{ll}
\textstyle{D_{z}=\frac{1}{\sqrt{2}}}(D_{x}-iD_{y})\,,~~~~&~~~~
\textstyle{D_{\bar{z}}=\frac{1}{\sqrt{2}}}(D_{x}+iD_{y})\,,\\
{}&{}\\
F_{za}=\textstyle{\frac{1}{\sqrt{2}}}(D_{z}X_{{2a}{-1}}-i D_{z}X_{{2a}})\,,~~~~&~~~~
F_{z\bar{a}}=\textstyle{\frac{1}{\sqrt{2}}}(D_{z}X_{{2\bar{a}}{-1}}+iD_{z}X_{{2\bar{a}}})\,.
\ea
\ee

%%%%%%%%%%%%%%%%%%%%%%%%%%%%%%%%%%%%%%%%%%%%%%%%%%%%%%%%%%%%%%%%%%%%%%%%%%%%%%%%%%%%%%%%%%%%%%%%%%%
%%%%%%%%%%%%%%%%%%%%%%%%%%%%%%%%%%%%%%%%%%%%%%%%%%%%%%%%%%%%%%%%%%%%%%%%%%%%%%%%%%%%%%%%%%%%%%%%%%%
\subsubsection{$N=4$ $\,\SO(2){\times\SU(2)}{\times\SO(4)}$ invariant BPS equations}
The $N=4$ $\,\SO(2){\times\SU(2)}{\times\SO(4)}$ invariant  BPS equations corresponding to the projection matrix (\ref{SO252}) are, up to Hermitian conjugation,
\be
\ba{lllll}
F_{z\bar{a}}=0\,,~~~&~~~F_{zp}=0\,,~~~&~~~F_{pab}=0\,,~~~&~~~F_{tI}-iF_{Ia}{}^{a}=0\,,~~~&~~~
F_{Ipq}+\half\epsilon_{pqrs\,}F_{I}{}^{rs}=0\,,
\ea
\ee
 where $I=1,2,\cdots,8$,  $~p,q,r,s=5,6,7,8$, $\epsilon_{pqrs}$ is a totally anti-symmetric tensor with  $\epsilon_{5678}{=1}$ and $a,b,\bar{a}=1,2$ such that  the $\SO(4)\subset\SO(8)$ coordinates are complexified   by  the complex structure ${\Gamma_{12}}{+\Gamma_{34}}$.

%%%%%%%%%%%%%%%%%%%%%%%%%%%%%%%%%%%%%%%%%%%%%%%%%%%%%%%%%%%%%%%%%%%%%%%%%%%%%%%%%%%%%%%%%%%%%%%%%%%
%%%%%%%%%%%%%%%%%%%%%%%%%%%%%%%%%%%%%%%%%%%%%%%%%%%%%%%%%%%%%%%%%%%%%%%%%%%%%%%%%%%%%%%%%%%%%%%%%%%
\subsubsection{$N=6$ $\,\SO(2){\times\SO(2)}{\times\SU(3)}$ invariant BPS equations}
The $N=6$ $\,\SO(2){\times\SO(2)}{\times\SU(3)}$ invariant  BPS equations corresponding to the projection matrix (\ref{SO253}) are, up to Hermitian conjugation,
\be
\ba{llll}
F_{z\bar{\omega}}=0\,,~~~~&~~~~F_{za}=0\,,~~~~&~~~~F_{z\bar{a}}=0\,,~~~~&~~~~
F_{t\omega}-i\textstyle{\frac{1}{3}}F_{\omega a}{}^{a}=0\,,\\
{}&{}&{}&{}\\
F_{ta}-iF_{a\omega\bar{\omega}}=0\,,~~~~&~~~~
F_{\omega ab}=0\,,~~~~&~~~~F_{ab\bar{c}}=0\,,~~~~&~~~~F_{\omega a\bar{b}}-\textstyle{\frac{1}{3}}(F_{\omega c}{}^{c})\delta_{a\bar{b}}=0\,,
\ea
\ee
where   $a,\bar{a}=1,2,3$ such that we complexify  the $\SO(6)\subset\SO(8)$ coordinates  by  the complex structure ${\Gamma_{34}}{+\Gamma_{56}}{+\Gamma_{87}}$ and also set separately for $\SO(2)\subset\SO(8)$,
\be
\ba{ll}
F_{z\omega}\equiv\textstyle{\frac{1}{\sqrt{2}}}(F_{z1}-iF_{z2})\,,~~~~&~~~~
F_{z\bar{\omega}}\equiv\textstyle{\frac{1}{\sqrt{2}}}(F_{z1}+iF_{z2})\,.
\ea
\label{omegacom}
\ee

%%%%%%%%%%%%%%%%%%%%%%%%%%%%%%%%%%%%%%%%%%%%%%%%%%%%%%%%%%%%%%%%%%%%%%%%%%%%%%%%%%%%%%%%%%%%%%%%%%%
%%%%%%%%%%%%%%%%%%%%%%%%%%%%%%%%%%%%%%%%%%%%%%%%%%%%%%%%%%%%%%%%%%%%%%%%%%%%%%%%%%%%%%%%%%%%%%%%%%%
\subsubsection{$N=8$ $\,\SO(2){\times\SO(2)}{\times\SO(6)}$ invariant BPS equations}
The $N=8$ $\,\SO(2){\times\SO(2)}{\times\SO(6)}$ invariant BPS equations
corresponding to the projection matrix (\ref{SO254}) are, up to Hermitian conjugation,
\be
\ba{llll}
F_{z\bar{\omega}}=0\,,~~~~~&~~~~~F_{zp}=0\,,~~~~&~~~~
F_{tI}-iF_{I\omega\bar{\omega}}=0\,,~~~~&~~~~F_{Ipq}=0\,,
\ea
\label{SO2N8}
\ee
where $I=1,2,\cdots,8$, $~p=3,4,5,6,7,8$ and  we complexify  the $\SO(2)\subset\SO(8)$ coordinates  by  the complex structure ${\Gamma_{12}}$ to employ (\ref{omegacom}).

%%%%%%%%%%%%%%%%%%%%%%%%%%%%%%%%%%%%%%%%%%%%%%%%%%%%%%%%%%%%%%%%%%%%%%%%%%%%%%%%%%%%%%%%%%%%%%%%%%%
%%%%%%%%%%%%%%%%%%%%%%%%%%%%%%%%%%%%%%%%%%%%%%%%%%%%%%%%%%%%%%%%%%%%%%%%%%%%%%%%%%%%%%%%%%%%%%%%%%%
%%%%%%%%%%%%%%%%%%%%%%%%%%%%%%%%%%%%%%%%%%%%%%%%%%%%%%%%%%%%%%%%%%%%%%%%%%%%%%%%%%%%%%%%%%%%%%%%
%%%%%%%%%%%%%%%%%%%%%%%%%%%%%%%%%%%%%%%%%%%%%%%%%%%%%%%%%%%%%%%%%%%%%%%%%%%%%%%%%%%%%%%%%%%%%%%%
\subsection{$\SO(1,1)$ invariant BPS equations\label{secBPSSO11}}
The generic $N=1$  projection matrix~(\ref{SO11OG}) leads to the following
 $N{=1}$ $\,\SO(1,1){\times\SO(7)}$ invariant BPS equations
which involve   four free signs $\,\alpha_{0}^{2}=\alpha_{1}^{2}=\alpha_{2}^{2}=\alpha_{3}^{2}=1$:
\be
\ba{l}
F_{tI}-\alpha_{0}F_{xI}=0\,,~~~~~~~~~~~I=1,2,\cdots,8\,,\\
{}\\
\alpha_{0}F_{y1}-{\alpha_{1}\alpha_{2}}F_{278}-{\alpha_{2}\alpha_{3}}F_{548}-{\alpha_{3}\alpha_{1}}F_{638}
-\alpha_{1}F_{234}-\alpha_{2}F_{256}-\alpha_{3}F_{357}-{\alpha_{1}\alpha_{2}\alpha_{3}}F_{476}=0\,,\\
{}\\
\alpha_{0}F_{y2}-{\alpha_{1}\alpha_{2}}F_{718}-{\alpha_{2}\alpha_{3}}F_{376}-{\alpha_{3}\alpha_{1}}F_{475}
-\alpha_{1}F_{143}-\alpha_{2}F_{165}-\alpha_{3}F_{468}-{\alpha_{1}\alpha_{2}\alpha_{3}}F_{538}=0\,,\\
{}\\
\alpha_{0}F_{y3}-{\alpha_{1}\alpha_{2}}F_{456}-{\alpha_{2}\alpha_{3}}F_{267}-{\alpha_{3}\alpha_{1}}F_{168}
-\alpha_{1}F_{124}-\alpha_{2}F_{478}-\alpha_{3}F_{517}-{\alpha_{1}\alpha_{2}\alpha_{3}}F_{258}=0\,,\\
{}\\
\alpha_{0}F_{y4}-{\alpha_{1}\alpha_{2}}F_{536}-{\alpha_{2}\alpha_{3}}F_{158}-{\alpha_{3}\alpha_{1}}F_{257}
-\alpha_{1}F_{132}-\alpha_{2}F_{738}-\alpha_{3}F_{628}-{\alpha_{1}\alpha_{2}\alpha_{3}}F_{167}=0\,,\\
{}\\
\alpha_{0}F_{y5}-{\alpha_{1}\alpha_{2}}F_{346}-{\alpha_{2}\alpha_{3}}F_{418}-{\alpha_{3}\alpha_{1}}F_{427}
-\alpha_{1}F_{678}-\alpha_{2}F_{126}-\alpha_{3}F_{137}-{\alpha_{1}\alpha_{2}\alpha_{3}}F_{328}=0\,,\\
{}\\
\alpha_{0}F_{y6}-{\alpha_{1}\alpha_{2}}F_{354}-{\alpha_{2}\alpha_{3}}F_{273}-{\alpha_{3}\alpha_{1}}F_{318}
-\alpha_{1}F_{758}-\alpha_{2}F_{152}-\alpha_{3}F_{248}-{\alpha_{1}\alpha_{2}\alpha_{3}}F_{174}=0\,,\\
{}\\
\alpha_{0}F_{y7}-{\alpha_{1}\alpha_{2}}F_{128}-{\alpha_{2}\alpha_{3}}F_{236}-{\alpha_{3}\alpha_{1}}F_{245}
-\alpha_{1}F_{568}-\alpha_{2}F_{348}-\alpha_{3}F_{153}-{\alpha_{1}\alpha_{2}\alpha_{3}}F_{146}=0\,,\\
{}\\
\alpha_{0}F_{y8}+{\alpha_{1}\alpha_{2}}F_{127}+{\alpha_{2}\alpha_{3}}F_{154}+{\alpha_{3}\alpha_{1}}F_{163}
+\alpha_{1}F_{567}+\alpha_{2}F_{347}+\alpha_{3}F_{246}+{\alpha_{1}\alpha_{2}\alpha_{3}}F_{253}=0\,.
\ea
\label{SO11BPSeqMaster}
\ee
The above set of  BPS equations can be regarded as the master equations for generic
$\SO(1,1)$ invariant BPS equations.  One can classify the BPS equations according to the decomposition of the number of preserved supersymmetries as  $(N_{+},N_{-})$ (\ref{Npmdef}). Among others,  below we spell explicitly
$(N_{+},\,0\,)$  as well as  $(N,N)$  BPS equations with $N_{+}=1,2,\cdots,7$, $~N=1,2,3,4$.

%%%%%%%%%%%%%%%%%%%%%%%%%%%%%%%%%%%%%%%%%%%%%%%%%%%%%%%%%%%%%%%%%%%%%%%%%%%%%%%%%%%%%%%%%%%%%%%%%%%%%%%%%%%%%%%%
%%%%%%%%%%%%%%%%%%%%%%%%%%%%%%%%%%%%%%%%%%%%%%%%%%%%%%%%%%%%%%%%%%%%%%%%%%%%%%%%%%%%%%%%%%%%%%%%%%%%%%%%%%%%%%%%
\subsubsection{$(N_{+},N_{-})=(1,0)$  $\,\SO(1,1){\times\SO(7)}$ invariant  BPS equations - \textit{octonion}}
With the choice of ${(\alpha_{0},\alpha_{1},\alpha_{2},\alpha_{3})=(++++)}$,
the $(N_{+},N_{-}){=(1,0)}$ $\,\SO(1,1){\times\SO(7)}$ invariant BPS equations~(\ref{SO11BPSeqMaster}) assume a compact form:
\be
\ba{ll}
F_{tI}-F_{xI}=0\,,~~~~~&~~~~F_{y I}-\textstyle{\frac{1}{6}}\cC_{IJKL}F^{JKL}=0\,,
\ea
\label{SO11N10}
\ee
which generalizes the $N=2$  $\,\SO(1,2){\times\SO(7)}$ invariant  BPS equations~(\ref{SO12N2}).

%%%%%%%%%%%%%%%%%%%%%%%%%%%%%%%%%%%%%%%%%%%%%%%%%%%%%%%%%%%%%%%%%%%%%%%%%%%%%%%%%%%%%%%%%%%%%%%%
%%%%%%%%%%%%%%%%%%%%%%%%%%%%%%%%%%%%%%%%%%%%%%%%%%%%%%%%%%%%%%%%%%%%%%%%%%%%%%%%%%%%%%%%%%%%%%%%%%%%%%%%%%%%%%%%
%%%%%%%%%%%%%%%%%%%%%%%%%%%%%%%%%%%%%%%%%%%%%%%%%%%%%%%%%%%%%%%%%%%%%%%%%%%%%%%%%%%%%%%%%%%%%%%%%%%%%%%%%%%%%%%%
\subsubsection{$(N_{+},N_{-})=(2,0)$  $\,\SO(1,1){\times\SO(2)}{\times\SO(6)}$ invariant  BPS equations - \textit{complex}}
The $(N_{+},N_{-})=(2,0)$ $\,\SO(1,1){\times\SO(2)}{\times\SO(6)}$ invariant  BPS equations are, with $F_{tI}{-F_{xI}}{=0}$,
\be
\ba{ll}
\cJ^{IJ}F_{yJ}+\half F^{I}{}_{JK}\cJ^{JK}=0\,,~~~&~~~F_{IJK}=(1{\otimes\cJ}{\otimes\cJ}+\cJ{\otimes 1}{\otimes\cJ}+
\cJ{\otimes\cJ}{\otimes 1})_{IJK}{}^{LMN}F_{LMN}\,,
\ea
\label{SO11N20}
\ee
which generalizes the $N=4$  $\,\SO(1,2){\times\SO(6)}{\times\SO(2)}$ invariant  BPS equations~(\ref{SO12N4}).

%%%%%%%%%%%%%%%%%%%%%%%%%%%%%%%%%%%%%%%%%%%%%%%%%%%%%%%%%%%%%%%%%%%%%%%%%%%%%%%%%%%%%%%%%%%%%%%%%%%%%%%%%%%%%%%%
%%%%%%%%%%%%%%%%%%%%%%%%%%%%%%%%%%%%%%%%%%%%%%%%%%%%%%%%%%%%%%%%%%%%%%%%%%%%%%%%%%%%%%%%%%%%%%%%%%%%%%%%%%%%%%%%
\subsubsection{$(N_{+},N_{-})=(3,0)$  $\,{\SO(1,1)}{\times\SO(3)}{\times\SO(5)}$ invariant  BPS equations - \textit{quarternion}}
The $(N_{+},N_{-})=(3,0)$  $\,{\SO(1,1)}{\times\SO(3)}{\times\SO(5)}$ invariant  BPS equations are, with $F_{tI}{-F_{xI}}{=0}$,
\be
\cJ_{p}^{IJ}F_{yJ}+\half F^{I}{}_{JK}\cJ_{p}^{JK}=0\,,~~~~~~~~p=1,2,3\,,
\label{SO11N30}
\ee
where $\cJ_{1},\cJ_{2},\cJ_{3}$ are three distinct complex structures satisfying the quaternion relations,
$\cJ_{1}^{2}=\cJ_{2}^{2}=\cJ_{3}^{2}=\cJ_{1}\cJ_{2}\cJ_{3}=-1$ (\ref{J3}).
It is worth to note that the remaining relation  of (\ref{SO11N20})
$F_{(3,0)}=0$  is fulfilled   automatically for each complex structure. Eq.(\ref{SO11N30}) generalizes the $N=6$  $\,\SO(1,2){\times\SO(5)}{\times\SO(3)}$ invariant  BPS equations~(\ref{SO12N6}).

%%%%%%%%%%%%%%%%%%%%%%%%%%%%%%%%%%%%%%%%%%%%%%%%%%%%%%%%%%%%%%%%%%%%%%%%%%%%%%%%%%%%%%%%%%%%%%%%%%%%%%%%%%%%%%%%
%%%%%%%%%%%%%%%%%%%%%%%%%%%%%%%%%%%%%%%%%%%%%%%%%%%%%%%%%%%%%%%%%%%%%%%%%%%%%%%%%%%%%%%%%%%%%%%%%%%%%%%%%%%%%%%%
\subsubsection{$(N_{+},N_{-})=(4,0)$  $\,{\SO(1,1)}{\times\SO(4)}{\times\SO(4)}$ invariant  BPS equations}
The $(N_{+},N_{-})=(4,0)$  $\,{\SO(1,1)}{\times\SO(4)}{\times\SO(4)}$ invariant  BPS equations are, with $F_{tI}{-F_{xI}}{=0}$,
\be
\cT_{IJKL}F_{y}{}^{L}+F_{IJK}+\half F_{I}{}^{LM}\cT_{JKLM}+\half F_{J}{}^{LM}\cT_{KILM}+\half F_{K}{}^{LM}\cT_{IJLM}=0\,,
\label{SO11N40}
\ee
where $\cT_{IJKL}$ is a ${\SO(4)}{\times\SO(4)}$ invariant self-dual four-form (\ref{Tsdff}). Eq.(\ref{SO11N40})  generalizes the $N=8$  $\,\SO(1,2){\times\SO(4)}{\times\SO(4)}$ invariant  BPS equations~(\ref{SO12N8}). Some mass deformations of the above  BPS equations are studied in Ref.\cite{Hosomichi:2008qk}.

%%%%%%%%%%%%%%%%%%%%%%%%%%%%%%%%%%%%%%%%%%%%%%%%%%%%%%%%%%%%%%%%%%%%%%%%%%%%%%%%%%%%%%%%%%%%%%%%%%%%%%%%%%%%%%%%
%%%%%%%%%%%%%%%%%%%%%%%%%%%%%%%%%%%%%%%%%%%%%%%%%%%%%%%%%%%%%%%%%%%%%%%%%%%%%%%%%%%%%%%%%%%%%%%%%%%%%%%%%%%%%%%%
\subsubsection{$(N_{+},N_{-})=(5,0)$   $\,{\SO(1,1)}{\times\SO(5)}{\times\SO(3)}$ invariant  BPS equations}
  The $(N_{+},N_{-})=(5,0)$    $\,{\SO(1,1)}{\times\SO(5)}{\times\SO(3)}$ invariant  BPS equations are,
  with $F_{tI}{-F_{xI}}{=0}$,
\be
\ba{ll}
\cT_{IJKL}F_{y}{}^{L}+F_{IJK}+\textstyle{\frac{3}{2}}F_{[I}{}^{LM}\cT_{JK]LM}=0\,,~~~~~&~~~~~
\cJ^{IJ}F_{yJ}+\half F_{IJK}\cJ^{JK}=0\,,
\label{SO11N50}
\ea
\ee
where $\cT_{IJKL}$ and $\cJ^{IJ}$ are given in (\ref{TJ}).  Eq.(\ref{SO11N50}) generalizes the $N=10$  $\,\SO(1,2){\times\SO(3)}{\times\SO(5)}$ invariant  BPS equations~(\ref{SO12N10}).

%%%%%%%%%%%%%%%%%%%%%%%%%%%%%%%%%%%%%%%%%%%%%%%%%%%%%%%%%%%%%%%%%%%%%%%%%%%%%%%%%%%%%%%%%%%%%%%%%%%%%%%%%%%%%%%%
%%%%%%%%%%%%%%%%%%%%%%%%%%%%%%%%%%%%%%%%%%%%%%%%%%%%%%%%%%%%%%%%%%%%%%%%%%%%%%%%%%%%%%%%%%%%%%%%%%%%%%%%%%%%%%%%
\subsubsection{$(N_{+},N_{-})=(6,0)$    $\,{\SO(1,1)}{\times\SO(6)}{\times\SO(2)}$ invariant  BPS equations}
The $(N_{+},N_{-})=(6,0)$   $\,{\SO(1,1)}{\times\SO(6)}{\times\SO(2)}$ invariant  BPS equations are,  $F_{tI}{-F_{xI}}{=0}$,
\be
\cT_{p}^{IJ}F_{yJ}+\half F^{I}{}_{JK}\cT_{p}^{JK}=0\,,~~~~~~~~p=1,2,3,4,5,6\,,
\label{SO11N60}
\ee
where six of two-forms $\cT_{p}$, $p=1,2,\cdots,6$ are given in (\ref{T6def}).
Eq.(\ref{SO11N60})  generalizes the $N=12$  $\,\SO(1,2){\times\SO(2)}{\times\SO(6)}$ invariant  BPS equations~(\ref{SO12N12}).

%%%%%%%%%%%%%%%%%%%%%%%%%%%%%%%%%%%%%%%%%%%%%%%%%%%%%%%%%%%%%%%%%%%%%%%%%%%%%%%%%%%%%%%%%%%%%%%%%%%%%%%%%%%%%%%%
%%%%%%%%%%%%%%%%%%%%%%%%%%%%%%%%%%%%%%%%%%%%%%%%%%%%%%%%%%%%%%%%%%%%%%%%%%%%%%%%%%%%%%%%%%%%%%%%%%%%%%%%%%%%%%%%
\subsubsection{$(N_{+},N_{-})=(7,0)$    $\,{\SO(1,1)}{\times\SO(7)}$ invariant  BPS equations}
The $(N_{+},N_{-})=(7,0)$   $\,{\SO(1,1)}{\times\SO(7)}$ invariant  BPS equations are,  with $F_{tI}{-F_{xI}}{=0}$,
\be
\cT_{p}^{IJ}F_{yJ}+\half F^{I}{}_{JK}\cT_{p}^{JK}=0\,,~~~~~~~~p=1,2,3,4,5,6,7\,.
\label{SO11N70}
\ee
Here we have seven of two-forms, six given by (\ref{T6def}) and  last one by
\be
\half\cT_{7}^{IJ}\Gamma_{IJ}=\Gamma^{13}+\Gamma^{57}\,.
\ee
They form a fundamental representation of $\SO(7)$.\\

%%%%%%%%%%%%%%%%%%%%%%%%%%%%%%%%%%%%%%%%%%%%%%%%%%%%%%%%%%%%%%%%%%%%%%%%%%%%%%%%%%%%%%%%%%%%%%%%%%%%%%%%%%%%%%%%
%%%%%%%%%%%%%%%%%%%%%%%%%%%%%%%%%%%%%%%%%%%%%%%%%%%%%%%%%%%%%%%%%%%%%%%%%%%%%%%%%%%%%%%%%%%%%%%%%%%%%%%%%%%%%%%%
\subsubsection{$(N_{+},N_{-})=(1,1)$    $\,{\SO(1,1)}{\times\SO(6)}$ invariant  BPS equations}
The $(N_{+},N_{-})=(1,1)$   $\,{\SO(1,1)}{\times\SO(6)}$ invariant  BPS equations are,  with $F_{tI}=F_{xI}=0$, best expressed in  complex coordinates,
\be
\ba{ll}
F_{ab}{}^{b}=0\,,~~~~~&~~~~~F_{y\bar{a}}-\textstyle{\frac{1}{3}}\epsilon_{\bar{a}}{}^{bcd}F_{bcd}=0\,.
\label{SO11N11}
\ea
\ee

%%%%%%%%%%%%%%%%%%%%%%%%%%%%%%%%%%%%%%%%%%%%%%%%%%%%%%%%%%%%%%%%%%%%%%%%%%%%%%%%%%%%%%%%%%%%%%%%%%%%%%%%%%%%%%%%
%%%%%%%%%%%%%%%%%%%%%%%%%%%%%%%%%%%%%%%%%%%%%%%%%%%%%%%%%%%%%%%%%%%%%%%%%%%%%%%%%%%%%%%%%%%%%%%%%%%%%%%%%%%%%%%%
\subsubsection{$(N_{+},N_{-})=(2,2)$    $\,{\SO(1,1)}{\times\SO(2)}{\times\SO(2)}{\times\SO(4)}$ invariant  BPS equations}
The $(N_{+},N_{-})=(2,2)$   $\,{\SO(1,1)}{\times\SO(2)}{\times\SO(2)}{\times\SO(4)}$ invariant  BPS equations are,  with $F_{tI}=F_{xI}=0$,
\be
(3\cJ^{[IJ}\cJ^{K]L}-\cT^{IJKL})F_{yL}+F^{IJK}+\textstyle{\frac{3}{2}}F^{[I}{}_{LM}\cT^{JK]LM}=0\,,
\label{SO11N22}
\ee
where $\cJ^{IJ}$ is  the  complex structure of $\Gamma^{12}+\Gamma^{34}+\Gamma^{56}+\Gamma^{78}$ (\ref{cJrep})  and
$\cT^{IJKL}$ is the  self-dual ${\SO(4)}{\times\SO(4)}$ invariant four-form tensor of $\Gamma^{1234}+\Gamma^{5678}$ (\ref{Tsdff}).\\

%%%%%%%%%%%%%%%%%%%%%%%%%%%%%%%%%%%%%%%%%%%%%%%%%%%%%%%%%%%%%%%%%%%%%%%%%%%%%%%%%%%%%%%%%%%%%%%%%%%%%%%%%%%%%%%%
%%%%%%%%%%%%%%%%%%%%%%%%%%%%%%%%%%%%%%%%%%%%%%%%%%%%%%%%%%%%%%%%%%%%%%%%%%%%%%%%%%%%%%%%%%%%%%%%%%%%%%%%%%%%%%%%
\subsubsection{$(N_{+},N_{-})=(3,3)$    $\,{\SO(1,1)}{\times\SO(3)}{\times\SO(3)}{\times\SO(2)}$ invariant  BPS equations}
We present the $(N_{+},N_{-})=(3,3)$   $\,{\SO(1,1)}{\times\SO(3)}{\times\SO(3)}{\times\SO(2)}$ invariant  BPS equations with a pair of quarternion  structures, one    from (\ref{J3}) and the other  from
$\Gamma^{12}{+\Gamma^{87}}{+\Gamma^{56}}{+\Gamma^{43}}$,
$~\Gamma^{17}{+\Gamma^{28}}{+\Gamma^{53}}{+\Gamma^{64}}$,
$~\Gamma^{18}{+\Gamma^{72}}{+\Gamma^{54}}{+\Gamma^{36}}$. With $F_{tI}=F_{xI}=0$ they are
\be
\ba{ll}
\cJ_{p}^{IJ}F_{yJ}+\half F^{I}{}_{JK}\cJ_{p}^{JK}=0\,,~~~~~&~~~~~\hat{\cJ}_{p}^{IJ}F_{yJ}-\half F^{I}{}_{JK}\hat{\cJ}_{p}^{JK}=0\,,~~~~~~~~~~p=1,2,3\,.
\label{SO11N33}
\ea
\ee
{}\\

%%%%%%%%%%%%%%%%%%%%%%%%%%%%%%%%%%%%%%%%%%%%%%%%%%%%%%%%%%%%%%%%%%%%%%%%%%%%%%%%%%%%%%%%%%%%%%%%%%%%%%%%%%%%%%%%
%%%%%%%%%%%%%%%%%%%%%%%%%%%%%%%%%%%%%%%%%%%%%%%%%%%%%%%%%%%%%%%%%%%%%%%%%%%%%%%%%%%%%%%%%%%%%%%%%%%%%%%%%%%%%%%%
\subsubsection{$(N_{+},N_{-})=(4,4)$    $\,{\SO(1,1)}{\times\SO(4)}{\times\SO(4)}$ invariant  BPS equations}
The $(N_{+},N_{-})=(4,4)$    $\,{\SO(1,1)}{\times\SO(4)}{\times\SO(4)}$ invariant  BPS equations are,  with $F_{tI}=F_{xI}=0$ , in terms of the self-dual ${\times\SO(4)}{\times\SO(4)}$ invariant four-form tensor,
\be
\cT^{IJKL}F_{yL}+F^{IJK}=0\,.
\label{SO11N44}
\ee
Especially among all the half BPS cases \textit{i.e.} $N_{+}+N_{-}=8$,   only the case  $(N_{+},N_{-})=(4,4)$  leads to the  nontrivial BPS equations. \\
%%%%%%%%%%%%%%%%%%%%%%%%%%%%%%%%%%%%%%%%%%%%%%%%%%%%%%%%%%%%%%%%%%%%%%%%%%%%%%%%%%%%%%%%%%%%%%%%%%%%%%%%%%%%%%%%
%%%%%%%%%%%%%%%%%%%%%%%%%%%%%%%%%%%%%%%%%%%%%%%%%%%%%%%%%%%%%%%%%%%%%%%%%%%%%%%%%%%%%%%%%%%%%%%%%%%%%%%%%%%%%%%%

\section{Discussion\label{discussion}}
In this paper we studied and identified a number of BPS equations for the multiple M2-brane
theory proposed recently by Bagger and Lambert. We employed a method which had been
successfully applied to several analogous problems. One first constructs the basic projection matrices for the supersymmetry parameters, and then obtain the corresponding  BPS equations. Our classifications are complete for $\SO(1,2)$ as well as $\SO(2)^{\mathbf{5}}$ invariant BPS equations, while may be not for $\SO(1,1)$ invariant cases.

The BPS equations with different types and numbers of preserved supersymmetries are derived
in terms of the associated tensors which are invariant under the symmetry group
of the relevant BPS equations. In particular we derived three types of half BPS equations, which we recall:
\begin{itemize}
\item $N{=8}$  $\,{\SO(1,2)}{\times\SO(4)}{\times\SO(4)}$ invariant  BPS equations (\ref{SO12N8})
\be
\ba{ll}
F_{\mu I}=0\,,~~~~&~~~~
F_{IJK}+\half F_{I}{}^{LM}\cT_{JKLM}+\half F_{J}{}^{LM}\cT_{KILM}+\half F_{K}{}^{LM}\cT_{IJLM}=0\,.
\ea
\ee

\item  $N=8$ $\,\SO(2){\times\SO(2)}{\times\SO(6)}$ invariant BPS equations (\ref{SO2N8})
\be
\ba{llll}
F_{z\bar{\omega}}=0\,,~~~~~&~~~~~F_{zp}=0\,,~~~~&~~~~
F_{tI}-iF_{I\omega\bar{\omega}}=0\,,~~~~&~~~~F_{Ipq}=0\,,
\ea
\ee
where $I=1,2,\cdots,8$, $~p=3,4,5,6,7,8$,  and $\omega,\bar{\omega}$ are complex coordinates for   $\SO(2)\subset\SO(8)$.

\item  $(N_{+},N_{-})=(4,4)$    $\,{\SO(1,1)}{\times\SO(4)}{\times\SO(4)}$ invariant  BPS equations (\ref{SO11N44})
\be
\ba{ll}
F_{tI}=F_{xI}=0\,,~~~~~&~~~~~\cT^{IJKL}F_{yL}+F^{IJK}=0\,.
\ea
\ee

\end{itemize}

The BPS equations for different number of
supersymmetries exhibit the division algebra structures: octonion,
quarternion or complex. Let us take the Lorentz invariant type as examples.
For the least supersymmetric configurations preserving 1/8 supersymmetries, the relevant symmetry is $\SO(1,2){\times\SO(7)}$ and the
BPS equations can be elegantly written in terms of the invariant four-form which has close relation to octonions.
For 1/4-BPS
equations the symmetry is $\SO(1,2){\times\SO(6)}{\times\SO(2)}$ and  a complex structure appears. We next have 3/8 $\SO(1,2){\times\SO(5)}{\times\SO(3)}$ invariant BPS equations, which are
naturally best expressed in terms of quarternions or hyper-K\"ahler structure. In addition, for 1/2-BPS equations we have the $\SO(4)\times\SO(4)$ invariant self-dual four-form structure. We have also
identified the exotic classes with more than 1/2 supersymmetry. Apparently the governing
symmetries include more than one hyper-K\"ahler structures, but we have not been
able to express the BPS equations in a succinct way. The true mathematical identity
of such systems certainly deserves more careful study.

The explicit solutions of the BPS equations will give the spectrum of supersymmetric
solitons in Bagger-Lambert theory. It is natural to ask the $\cM$-theory interpretation
of such objects. The real scalar fields $X^{I}$ describe the locations of M2-branes in
the transverse $\mathbb{R}^{8}$. The spatial dependence of $X^{I}$ thus informs us on the shape
of M2-branes, or how they are embedded in the transverse
$\mathbb{R}^{8}$.  Eq.(\ref{SO25BPSeqMaster1}) and the subsequent analysis clearly suggest
that the M2-brane worldvolume should occupy holomorphic curves, which is natural for
supersymmetry. Likewise, time-dependence of the scalar field obviously implies that
there is momentum along the particular direction. The three-algebra terms $F_{IJK}$
describe the truly $\cM$-theoretic phenomena: polarization of multiple M2-branes into
M5-branes. Generically the BPS equations are given as various combinations of such basic
building blocks, and  more detailed descriptions with explicit solutions will be reported
in a separate publication.\\
{}\\
{}\\

%%%%%%%%%%%%%%%%%%%%%%%%%%%%%%%%%%%%%%%%%%%%%%%%%%%%%%%%%%%%%%%%%%%%%%%%%%%%%%%%%%%%%%%%%%%%%%%%%%
%%%%%%%%%%%%%%%%%%%%%%%%%%%%%%%%%%%%%%%%%%%%%%%%%%%%%%%%%%%%%%%%%%%%%%%%%%%%%%%%%%%%%%%%%%%%%%%%%%
%%%%%%%%%%%%%%%%%%%%%%%%%%%%%%%%%%%%%%%%%%%%%%%%%%%%%%%%%%%%%%%%%%%%%%%%%%%%%%%%%%%%%%%%%%%%%%%%%%
%%%%%%%%%%%%%%%%%%%%%%%%%%%%%%%%%%%%%%%%%%%%%%%%%%%%%%%%%%%%%%%%%%%%%%%%%%%%%%%%%%%%%%%%%%%%%%%%%%
%%\section{Solutions and comments\label{secsol}}
%%

%%%%%%%%%%%%%%%%%%%%%%%%%%%%%%%%%%%%%%%%%%%%%%%%%%%%%%%%%%%%%%%%%%%%%%%%%%%%%%%%%%%%%%%%%%%%%%%%
%%%%%%%%%%%%%%%%%%%%%%%%%%%%%%%%%%%%%%%%%%%%%%%%%%%%%%%%%%%%%%%%%%%%%%%%%%%%%%%%%%%%%%%%%%%%%%%%
\section*{Acknowledgments}
%%%%%%%%%%%%%%%%%%%%%%%%%%%%%%%%%%%%%%%%%%%%%%%%%%%%%%%%%%%%%%%%%%%%%%%%%%%%%%%%%%%%%%%%%%%%%%%%
%%%%%%%%%%%%%%%%%%%%%%%%%%%%%%%%%%%%%%%%%%%%%%%%%%%%%%%%%%%%%%%%%%%%%%%%%%%%%%%%%%%%%%%%%%%%%%%%
We wish to thank Bum-Hoon Lee for discussions and encouragement.
 This work is  supported by  the Center for Quantum Spacetime of Sogang
 University with grant number R11 - 2005 - 021.
 NK is partly supported by Korea Research Foundation Grant, No. KRF-2007-331-C00072.
The research  of JHP is supported in part by the Korea Science and Engineering Foundation grant funded by the Korea government (R01-2007-000-20062-0).
\newpage
\appendix

%%%%%%%%%%%%%%%%%%%%%%%%%%%%%%%%%%%%%%%%%%%%%%%%%%%%%%%%%%%%%%%%%%%%%%%%%%%%%%%%%%%%%%%%%%%%%%%%%%%%%%
%%%%%%%%%%%%%%%%%%%%%%%%%%%%%%%%%%%%%%%%%%%%%%%%%%%%%%%%%%%%%%%%%%%%%%%%%%%%%%%%%%%%%%%%%%%%%%%%%%%%%%
%%%%%%%%%%%%%%%%%%%%%%%%%%%%%%%%%%%%%%%%%%%%%%%%%%%%%%%%%%%%%%%%%%%%%%%%%%%%%%%%%%%%%%%%%%%%%%%%%%%%%%%%%%%
\section{Gamma matrices and octonions\label{Appoct}}
The eleven-dimensional $32{\times 32}$ gamma matrices $\Gamma^{M}$, $M=\mu,I$, $\mu=t,x,y$, $I=1,2,\cdots,8$ in  the  Bagger-Lambert theory naturally decompose into two parts: $\SO(1,2)$ the M2-brane worldvolume and  $\SO(8)$ the transverse space,
\be
\ba{llll}
\Gamma^{t}=\epsilon\otimes\gamma_{(9)}\,,~~&~~~
\Gamma^{x}=\sigma_{1}\otimes\gamma_{(9)}\,,~~&~~~
\Gamma^{y}=\sigma_{3}\otimes\gamma_{(9)}\,,~~&~~~
\Gamma^{I}=1\otimes\gamma^{I}\,,~~~~~~I=1,2,\cdots,8\,.
\ea
\label{txyI}
\ee
Here $\gamma^{I}$'s are the $16{\times 16}$  gamma matrices in the  eight-dimensional Euclidean space and $\gamma_{(9)}\equiv\gamma_{12\cdots8}$.  Clearly the $\SO(1,2)$  projection constraint  (\ref{chiral12}) coincides with that of $\SO(8)$,
\be
\Gamma^{txy}=1\otimes\gamma_{(9)}\,.
\ee
This  is consistent with the fact that the product of all the eleven-dimensional gamma matrices leads to the identity $\Gamma^{txy123\cdots 8}=1$. \\

Now we recall the seven quantities $\cE_{i}$, $i=1,2,3\cdots,7$  (\ref{Edef}). In the above choice of gamma matrices we have
\be
\ba{ll}
\cE_{i}=1\otimes E_{i}\,,~~~~~&~~~~~\cP=1\otimes P\,,
\ea
\ee
where as in (\ref{Edef})
\begin{equation}
\begin{array}{cccc}
E_{ 1}=\gamma_{8127}P\,,~&E_{ 2}=\gamma_{8163}P\,,~&E_{ 3}=\gamma_{8246}P\,,~& E_{4}=\gamma_{8347}P\,,\\
{}&{}&{}&{}\\
E_{ 5}=\gamma_{8567}P\,,~&E_{ 6}=\gamma_{8253}P\,,~&
E_{7}=\gamma_{8154}P\,,~&P=\half(1+\gamma_{(9)})\,.
\end{array}
\end{equation}
The  subscript spatial indices  of the  gamma matrices are organized such that the three  indices after the common $8$ are  identical to those of
the totally anti-symmetric octonionic  structure constants (\ref{octconst}).
%%%
%%\begin{equation}
%%\begin{array}{c}
%%~~e_{i}e_{j}=-\delta_{ij}+c_{ijk}\,e_{k}\,,~~~~~~~{i,j,k=1,2,\cdots,7\,,}\\ {}\\
%%1=c_{127}=c_{163}=c_{246}=c_{347}=c_{567}=c_{253}=c_{154}\,,~~~~~\mbox{others zero}\,.
%%\end{array}
%%\end{equation}
%%%
It is straightforward  to see that  $E_{i}$ forms a  representation of  the  ``square'' of the octonions on the eight-dimensional chiral space,
\begin{equation}
\begin{array}{cc}
E_{i}E_{j}=\delta_{ij}P+c^{\,2}_{ijk}\,E_{ k}\,,~~&~~
E_{ i}\equiv e_{i}\otimes e_{i}\,.
\end{array}
\label{square}
\end{equation}
Since they commute each other, they  form   a maximal set of the  mutually commuting  traceless  symmetric and real  matrices of the definite chirality
$\gamma_{(9)}E_{ i}=E_{i}$.  In fact, one can construct a $\SO(8)$  symmetric and real gamma matrix representation which makes all  $E_{i}$'s be  simultaneously diagonal, utilizing   the octonionic structure constants:
\be
\ba{lll}
\gamma_{\sI}=\left(\ba{cc}0~&~\rho_{\sI}\\{}&{}\\(\rho_{\sI})^{T}~&~0\ea\right)\,,~~&~~~
\rho_{\sI}(\rho_{\sJ})^{T}+\rho_{\sJ}(\rho_{\sI})^{T}=2{\delta_{\sI\sJ}}\,,
~~&~~~\gamma_{(9)}=\gamma_{12345678}=\left(\ba{rr}1&0\\0&-1\ea\right)\,.
\ea
\label{eightrep}
\ee
Here $\rho_{\sI}$,  $I=1,2,\cdots,8$ are $8{\times 8}$  real matrices given by\footnote{In particular, $\rho_{i}$, $1\leq i\leq 7$ correspond to the Majorana gamma matrices in Euclidean seven dimensions $\rho_{i}\rho_{j} +\rho_{j}\rho_{i} =-2\delta_{ij}$.}
\be
\ba{ll}
\rho_{i}=-(\rho_{i})^{T}=\left(\ba{cc}c_{i}~&~-n_{i}\\{}&{}\\
(n_{i})^{T}~&~0\ea\right),~~~i=1,2,\cdots,7\,,~~~~&~~~~~\rho_{8}=1\,,
\ea
\label{rhodef}
\ee
and $c_{i}$ is a  $7{\times 7}$ real matrix whose ${j,k}$ component is nothing but  the octonionic structure constant $c_{ijk}$ (\ref{octconst}), while $n_{i}$ is a seven-dimensional unit vector of which the $j$th component is  defined to be $\delta_{i}^{~j}$. \\

In the above choice of Majorana gamma matrix representation, all the $E_{i}$'s and $P$  are diagonal,
\be
\ba{l}
E_{1}=\diag(+1,+1,-1,-1,-1,-1,+1,+1,0,0,0,0,0,0,0,0)\,,\\
{}\\
E_{2}=\diag(+1,-1,+1,-1,-1,+1,-1,+1,0,0,0,0,0,0,0,0)\,,\\
{}\\
E_{3}=\diag(-1,+1,-1,+1,-1,+1,-1,+1,0,0,0,0,0,0,0,0)\,,\\
{}\\
E_{4}=\diag(-1,-1,+1,+1,-1,-1,+1,+1,0,0,0,0,0,0,0,0)\,,\\
{}\\
E_{5}=\diag(-1,-1,-1,-1,+1,+1,+1,+1,0,0,0,0,0,0,0,0)\,,\\
{}\\
E_{6}=\diag(-1,+1,+1,-1,+1,-1,-1,+1,0,0,0,0,0,0,0,0)\,,\\
{}\\
E_{7}=\diag(+1,-1,-1,+1,+1,-1,-1,+1,0,0,0,0,0,0,0,0)\,,\\
{}\\
P=\diag(+1,+1,+1,+1,+1,+1,+1,+1,0,0,0,0,0,0,0,0)\,,
\ea
\label{Es}
\ee
and the $\mbox{SO}(8)$ triality among ${\bf 8}_{{\rm v}}$, ${\bf 8}_{+}$, ${\bf 8}_{-}$  is apparent as  the ${\bf 8}_{{\rm v}}$ generators decompose into the   ${\bf 8}_{+}$ and ${\bf 8}_{-}$ generators,
\begin{equation}
\gamma_{IJ}=\left(\begin{array}{cc}\rho_{[I}\rho^{T}_{J]}&0\\ 0&\rho_{[I}^{T}\rho_{J]}\end{array}\right)\,.
\label{triality}
\end{equation}
With the identity $e_{8}\equiv 1$,
%%%
%%and $\rho_{\sI\sJ\sK}\equiv(\rho_{\sI})_{\sJ\sK}$,
%%%
the octonion algebra now spells completely:
\begin{equation}
\ba{ll}
e_{\sI}e_{\sJ}=(\rho_{\sI})_{\sJ\sK}\,e_{\sK}\,,~~~~&~~~~{I,J,K=1,2,\cdots,8\,.}
\ea
\end{equation}

Finally let us  consider a self-dual four-form and contract it with the  $\SO(8)$ gamma matrices $\Gamma^{IJKL}$, such as $\Upsilon_{4}\cP$ in (\ref{GOME}). Clearly utilizing the $\SO(8)$ triality, one can diagonalize $\Upsilon_{4}\cP$ to   express it as a linear combination of $\cE_{i}$'s.
This  shows that the canonical form of a self-dual four-form in eight dimensions  indeed takes the form  (\ref{canon4}): namely   the non-vanishing independent components are only those seven which are contracted to $\cE_{i}$'s.

%%%%%%%%%%%%%%%%%%%%%%%%%%%%%%%%%%%%%%%%%%%%%%%%%%%%%%%%%%%%%%%%%%%%%%%%%%%%%%%%%%%%%%%%%%%%%%%%
%%%%%%%%%%%%%%%%%%%%%%%%%%%%%%%%%%%%%%%%%%%%%%%%%%%%%%%%%%%%%%%%%%%%%%%%%%%%%%%%%%%%%%%%%%%%%%%%%%%%%%%%%%%%%%%%%%%%%%%%%%%%%
%%%%%%%%%%%%%%%%%%%%%%%%%%%%%%%%%%%%%%%%%%%%%%%%%%%%%%%%%%%%%%%%%%%%%%%%%%%%
\section{$\SO(2)$ invariant projection matrix\label{AppCartan}}
Here we derive the most general form of the $32\times 32$ projection matrices $\Omega$ which are  invariant under the Cartan subalgebra $\SO(2)^{{\mathbf{5}}}$  of $\SO(10)$, satisfying in addition to the conditions (\ref{Omegacon}),
\be
\ba{lllll}
[\Gamma^{xy},\Omega]=0\,,~~~&~~~[\Gamma^{12},\Omega]=0\,,~~~&~~~[\Gamma^{34},\Omega]=0\,,~~~&~~~
[\Gamma^{56},\Omega]=0\,,~~~&~~~[\Gamma^{78},\Omega]=0\,.
\ea
\ee
As  (\ref{SO25Omega}),  they  assume the general form:
\be
\Omega=\left[c+\Gamma^{xy}\!\left(a_{1}\Gamma^{12}+a_{2}\Gamma^{34}+a_{3}\Gamma^{56}+a_{4}\Gamma^{78}\right)+
b_{1}\Gamma^{1234}+b_{2}\Gamma^{1256}+b_{3}\Gamma^{1278}\right]\cP\,,
\ee
where $c,a_{1},\cdots,b_{3}$ are eight \textit{a priori}  unknown real constants which must be determined by requiring  the remaining condition  $\Omega^{2}=\Omega$. In particular the number of the preserved supersymmetries is related to the constant $c$ by
\be
N=\Tr\Omega=16 c\,.
\label{Nc}
\ee

It is convenient to reparameterize the four constants $a_{1}, a_{2}, a_{3}, a_{4}$ by four other constants $e_{1}, e_{2}, e_{3}, e_{4}$
\be
\ba{ll}
e_{1}=2(a_{1}+a_{2}+a_{3}+a_{4})\,,~~~~&~~~~
e_{2}=2(a_{1}+a_{2}-a_{3}-a_{4})\,,
\\
{}&{}\\
e_{3}=2(a_{1}-a_{2}+a_{3}-a_{4})\,,~~~~&~~~~
e_{4}=2(-a_{1}+a_{2}+a_{3}-a_{4})\,,
\ea
\ee
and the other four  constants $c, b_{1}, b_{2}, b_{3}$ by another set of four constants $f_{1}, f_{2}, f_{3}, f_{4}$
\be
\ba{ll}
f_{1}=2c-1-2b_{1}-2b_{2}-2b_{3}\,,~~~~&~~~~f_{2}=2c-1-2b_{1}+2b_{2}+2b_{3}\,,\\
{}&{}\\
f_{3}=2c-1+2b_{1}-2b_{2}+2b_{3}\,,~~~~&~~~~f_{4}=2c-1+2b_{1}+2b_{2}-2b_{3}\,.
\ea
\ee
It follows that
\be
\ba{ll}
a_{1}=\textstyle{\frac{1}{8}}(e_{1}+e_{2}+e_{3}-e_{4})\,,~~~~&~~~~
a_{2}=\textstyle{\frac{1}{8}}(e_{1}+e_{2}-e_{3}+e_{4})\,,\\
{}&{}\\
a_{3}=\textstyle{\frac{1}{8}}(e_{1}-e_{2}+e_{3}+e_{4})\,,~~~~&~~~~
a_{4}=\textstyle{\frac{1}{8}}(e_{1}-e_{2}-e_{3}-e_{4})\,,\\
{}&{}\\
b_{1}=\textstyle{\frac{1}{8}}(-f_{1}-f_{2}+f_{3}+f_{4})\,,~~~~&~~~~
b_{2}=\textstyle{\frac{1}{8}}(-f_{1}+f_{2}-f_{3}+f_{4})\,,\\
{}&{}\\
b_{3}=\textstyle{\frac{1}{8}}(-f_{1}+f_{2}+f_{3}-f_{4})\,,~~~~&~~~~
c=\textstyle{\frac{1}{8}}(f_{1}+f_{2}+f_{3}+f_{4}+4)\,.
\ea
\label{abcex}
\ee
Straightforward  calculation shows that $\Omega^{2}=\Omega$ is equivalent  for each $a=1,2,3,4$  to
\be
\ba{lll}
f_{a}e_{a}=0\,,~~~~&~~~~~e_{a}^{2}=(1+f_{a})(1-f_{a})~~~~~&~~~~~\mbox{not~~} a \mbox{~~sum}\,.
\ea
\ee
Hence for each $a$ we have four possible solutions:
\be
\ba{llll}
e_{a}=0\,,~\,f_{a}=+1\,;~~~&~~~e_{a}=0\,,~\,f_{a}=-1\,;~~~&~~~
e_{a}=+1\,,~\,f_{a}=0\,;~~~&~~~e_{a}=-1\,,~f_{a}=0\,.
\ea
\ee
Consequently from  (\ref{Nc}) and (\ref{abcex}),  the possible values of $c$ are $0$, $\frac{1}{8}$,  $\frac{2}{8}$,  $\frac{3}{8}$,  $\frac{4}{8}$,  $\frac{5}{8}$,  $\frac{6}{8}$,  $\frac{7}{8}$,  $1$, so that  the number of the preserved supersymmetries $N$ is an even number between zero and sixteen. The basic building blocks of all the possible projection matrices are those of $N=2$ given by
\be
\ba{ll}
\Omega&=\textstyle{\frac{1}{8}}
\left[1+\Gamma^{xy}\!\left(\beta_{1}\Gamma^{12}+\beta_{2}\Gamma^{34}+\beta_{3}\Gamma^{56}+
\beta_{1}\beta_{2}\beta_{3}\Gamma^{78}\right)-\beta_{1}\beta_{2}\Gamma^{1234}
-\beta_{3}\beta_{1}\Gamma^{1256}-\beta_{2}\beta_{3}\Gamma^{1278}\right]\!\cP\\
{}&{}\\
{}&=\textstyle{\frac{1}{8}}(1+\beta_{1}\Gamma^{xy12})(1+\beta_{2}\Gamma^{xy34})(1+\beta_{3}\Gamma^{xy56})\cP\,,
\ea
\ee
where $\beta_{1}$, $\beta_{2}$, $\beta_{3}$ are three independent signs,
\be
\beta_{1}^{2}=\beta_{2}^{2}=\beta_{3}^{2}=1\,.
\ee
There are eight possible $N=2$ projection matrices which are orthogonal to each other. By summing $k$ of them,   all the other generic projection matrices preserving $N=2k$ supersymmetries can be obtained.

%%
%%
%%Some useful relations are
%%\be
%%\gamma_{lmn}\gamma^{ijk}c_{ijk}=\gamma_{lmn}{}^{ijk}c_{ijk}+9\gamma_{[lm}{}^{ij}c_{n]ij}
%%-18\gamma_{[l}{}^{i}c_{mn]i}-6c_{lmn}\,.
%%\ee
%%

%%%%%%%%%%%%%%%%%%%%%%%%%%%%%%%%%%%%%%%%%%%%%%%%%%%%%%%%%%%%%%%%%%%%%%%%%%%%%%%%%%%%%%%%%%%%%%%%%%%%%%%%%%%%%%%%%%%%%%%%%%%%%
%%%%%%%%%%%%%%%%%%%%%%%%%%%%%%%%%%%%%%%%%%%%%%%%%%%%%%%%%%%%%%%%%%%%%%%%%%%%%%%%%%%%%%%%%%%%%%%%%%%%%%%%%%%%%%%%%%%%%%%%%%%%%

\end{document}